\documentclass[manuscript,nonacm]{acmart}

\usepackage[utf8]{inputenc}
\usepackage{subcaption}
\usepackage{mathtools} 
\usepackage{fancyvrb}
\usepackage{listings}
\usepackage{enumitem}
\usepackage{booktabs}
\usepackage{threeparttable}
\usepackage{textcomp}
\usepackage{placeins} 
\usepackage{cuted}    
\usepackage{capt-of}  

\usepackage[toc,page]{appendix}

\usepackage{amssymb}

\renewcommand\footnotetextcopyrightpermission[1]{} 
\AtBeginDocument{%
  
}

\begin{document}

\title{AnimAgents: Coordinating Multi-Stage Animation Pre-Production with Human–Multi-Agent Collaboration}


\author{Wen-Fan Wang}
\email{vann@cmlab.csie.ntu.edu.tw}
\orcid{0009-0001-1050-1170}
\affiliation{%
  \institution{National Taiwan University}
  \city{Taipei}
  \country{Taiwan}
}

\author{Chien-Ting Lu}
\authornote{These authors contributed equally as co-second authors.}
\email{B09902109@csie.ntu.edu.tw}
\orcid{0009-0000-8863-1277}
\affiliation{%
  \institution{National Taiwan University}
  \city{Taipei}
  \country{Taiwan}
}

\author{Jin Ping Ng}
\authornotemark[1]
\email{R12944057@csie.ntu.edu.tw}
\orcid{0009-0006-0999-5181}
\affiliation{%
  \institution{National Taiwan University}
  \city{Taipei}
  \country{Taiwan}
}

\author{Yi-Ting Chiu}
\authornotemark[1]
\email{R13922018@csie.ntu.edu.tw}
\orcid{0009-0004-0937-2705}
\affiliation{%
  \institution{National Taiwan University}
  \city{Taipei}
  \country{Taiwan}
}

\author{Ting-Ying Lee}
\email{tylee@cmlab.csie.ntu.edu.tw}
\orcid{0009-0002-4745-9460}
\affiliation{%
  \institution{National Taiwan University}
  \city{Taipei}
  \country{Taiwan}
}

\author{Miaosen Wang}
\email{wangmiaosen@gmail.com}
\orcid{0009-0004-1581-6552}
\affiliation{%
  \institution{Google DeepMind}
  \city{Mountain View, California}
  \country{USA}
}

\author{Bing-Yu Chen}
\email{robin@ntu.edu.tw}
\orcid{0000-0003-0169-7682}
\affiliation{%
  \institution{National Taiwan University}
  \city{Taipei}
  \country{Taiwan}
}

\author{Xiang 'Anthony' Chen}
\email{xac@ucla.edu}
\orcid{0000-0002-8527-1744}
\affiliation{%
  \institution{HCI Research, UCLA}
  \city{Los Angeles, California}
  \country{USA}
}

\renewcommand{\shortauthors}{Wang Lu Ng Chiu Lee Wang Chen Chen}


\begin{abstract}
Animation pre-production lays the foundation of an animated film by transforming initial concepts into a coherent blueprint across interdependent stages such as ideation, scripting, design, and storyboarding. While generative AI tools are increasingly adopted in this process, they remain isolated, requiring creators to juggle multiple systems without integrated workflow support. 
Our formative study with 12 professional creative directors and independent animators revealed key challenges in their current practice:
Creators must manually coordinate fragmented outputs, manage large volumes of information, and struggle to maintain continuity and creative control between stages. Based on the insights, we present AnimAgents, a human–multi-agent collaborative system that coordinates complex, multi-stage workflows through a core agent and specialized agents, supported by dedicated boards for the four major stages of pre-production.
AnimAgents enables stage-aware orchestration, stage-specific output management, and element-level refinement, providing an end-to-end workflow tailored to professional practice.
In a within-subjects summative study with 16 professional creators, AnimAgents significantly outperformed a strong single-agent baseline that equipped with advanced parallel image generation in coordination, consistency, information management, and overall satisfaction (p < .01). A field deployment with 4 creators further demonstrated AnimAgents' effectiveness in real-world projects.

\end{abstract}



\begin{teaserfigure}
  \includegraphics[width=\textwidth]{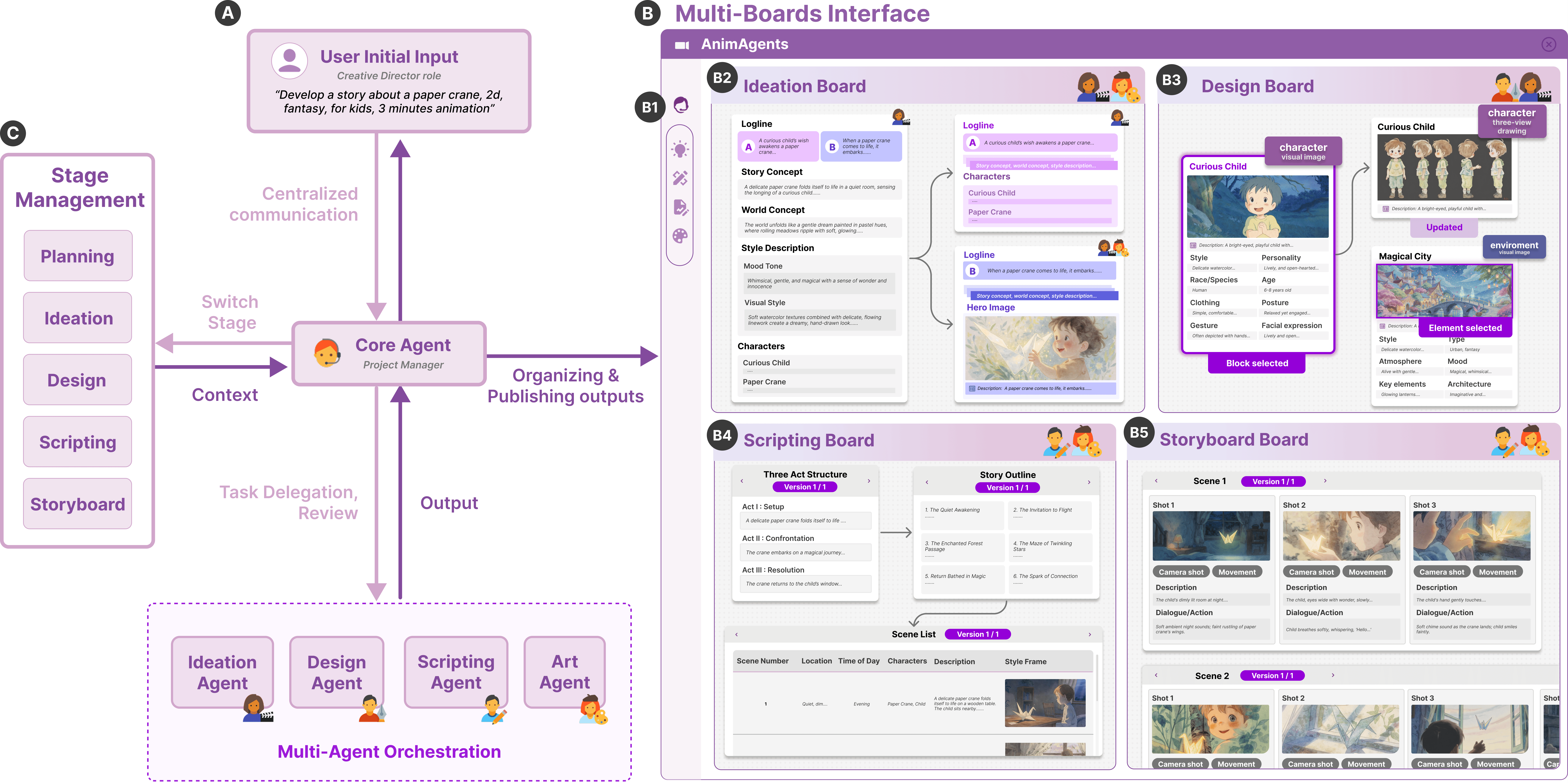}
  \caption{AnimAgents, a human–multi-agent collaborative system for animation pre-production. (A) Users provide an initial project input. (B) Outputs are organized into stage-specific boards—(B2) Ideation, (B3) Design, (B4) Scripting, and (B5) Storyboard—each tailored to its respective stage. Within these boards, a block- and element-based interaction model allows users to select specific items (e.g., the “Curious Child” block in (B3)) and delegate refinement or expansion to agents. (C) Stage-aware orchestration is handled by the Core Agent, which adapts its role to the active stage and coordinates four Specialized Agents (Ideation, Scripting, Design, and Art) to maintain coherence and continuity across the workflow.}
  \Description[This figure shows the architecture and interface of AnimAgents, where a Core Agent manages stages and specialized agents, and outputs are organized into ideation, design, scripting, and storyboard boards.]{This figure shows the architecture and main interface of AnimAgents, a human–multi-agent collaborative system for animation pre-production.  
  On the left, Stage Management is illustrated as five stages: Planning, Ideation, Design, Scripting, and Storyboard. At the center, the Core Agent acts as a project manager, coordinating communication, switching between stages, managing context, delegating tasks, and publishing outputs. Below the Core Agent, four specialized agents are shown: Ideation, Design, Scripting, and Art, representing multi-agent orchestration.  
  At the top, users provide an initial input prompt, such as developing a story concept, which enters the workflow. On the right, the Multi-Boards Interface organizes outputs into four stage-specific boards. The Ideation Board contains loglines, story concepts, world concepts, and style descriptions. The Design Board shows character sheets, such as a three-view drawing of the “Curious Child,” and environment designs like the “Magical City.” The Scripting Board includes a three-act structure, story outlines, and a scene list with details of characters, locations, and style frames. The Storyboard Board presents sequential panels for multiple scenes, with information on camera shots, movement, description, and dialogue or action.  
  The diagram emphasizes a block- and element-based interaction model: users can select a specific element, such as a logline or visual asset, or an entire block, and request refinement or expansion. Stage-aware orchestration ensures that the Core Agent adapts its role to the active stage and maintains coherence and continuity across the workflow.}
  \label{fig:hero_image}
\end{teaserfigure}


\maketitle


\section{INTRODUCTION}
\label{sec:introduction}
The animation pre-production process establishes the foundation of an animated film by turning an initial concept into a blueprint that defines its story and visual direction~\cite{wright2013animation, rall2017animation}. It spans interdependent stages, including ideation, scripting, visual design, and storyboard, where each builds on prior results and shapes downstream decisions~\cite{levy2010animation, meroz2021animation, williams2012animator}. In studios, story, design, and art departments collaborate under a creative director~\cite{o2007openpipeline}, while independent animators manage all stages themselves~\cite{simon2013producing}. For both, maintaining consistency across stages is essential, yet coordination remains a major challenge~\cite{li2011novel, liu2010preliminary}.

With the rapid adoption of generative AI (GenAI), creators increasingly use these tools in pre-production~\cite{shi2023understanding, gunanto2025future, epstein2023art}. Multimodal large language models (MLLMs) support narrative ideation and scriptwriting~\cite{qin2024charactermeet, singh2023hide}, while text-to-image (T2I) models aid design and visual storytelling~\cite{wang2025aideation, long2024sketchar}. Creators employ them for concept development, pitching and client communication, some even integrating them into production~\cite{wang2025gentune, zhang2025generative}.
Prior research has explored how AI can improve creative workflows in areas such as storytelling~\cite{yan2023xcreation, antony2025id}, visual design~\cite{wang2025aideation, tang2024s}, animation~\cite{abootorabi2025generative}, and motion graphics~\cite{Tseng2024KeyframerEA, ma2025mover}, improving controllability, efficiency, and creativity.
However, these advances focus on isolated steps rather than integrated workflows.
Although some commercial systems generate final outputs, such as storyboards or animations, they provide limited control, making it difficult for professionals to refine the outputs~\cite{izani2024impact, jingyang2024research}. As a result, creators must manually coordinate fragmented outputs across stages, making it difficult to maintain coherence throughout the project.

Recent advances in multi-agent systems (MAS) show promise in orchestrating complex processes such as software pipelines~\cite{han2025multi} and scientific research~\cite{schmidgall2025agent}. MAS distribute tasks across agents with autonomous planning and decision-making, operating in parallel to process large volumes of information and execute complex workflows~\cite{dorri2018multi, xi2025rise}. This paradigm is well-suited to creative domains like animation pre-production, where workflows span interdependent stages and require coordination across diverse outputs. Yet in creative applications such as film~\cite{xu2024filmagent} and animation~\cite{shi2025animaker}, MAS remain primarily optimized for end-to-end automation, often neglecting human involvement across stages or refinement of intermediate outputs. Addressing this gap, we propose a domain-grounded human–multi-agent collaborative system that coordinates professional pre-production stages into a cohesive workflow.

To investigate the challenges of current pre-production, we conducted a formative study with 12 professional animation creators, including studio creative directors and independent animators. Our analysis revealed these recurring challenges: Participants described the overwhelming task of managing large volumes of visual and textual material across stages and the difficulty of clearly communicating this information to teams or clients. Although AI tools were often used, they produced excessive output that required extensive manual filtering. The abundance of results made consistency hard to maintain, and participants often lost track of earlier decisions, while fragmented GenAI tools across stages further undermined narrative and visual coherence. Meanwhile, creators expressed a desire for GenAI to serve not just as an idea generator but also as a coordinator, supporting project management, organizing outputs, and easing communication with teams and clients.

Based on these findings, we present AnimAgents (Fig.~\ref{fig:hero_image}), a human–multi-agent collaborative system that addresses the complexity, variability, and continuity of animation pre-production workflows by coordinating across stages; challenges that are difficult to manage with ad hoc LLM chaining or isolated tools. AnimAgents serves two main user groups: studio creative directors, who use GenAI for exploration, references, and pitching, and independent animators, who need end-to-end, element-level support.
The system adopts a centralized structure~\cite{tran2025multi} (Fig.\ref{fig:hero_image}-A), with a Core Agent acting as project manager and Specialized Agents for ideation, scripting, art, and design, modeled after studio departmental workflows\cite{rall2017animation}. Specialized Agents’ outputs are reviewed, organized, and returned to the user through the Core Agent. AnimAgents' stage-aware orchestration (Fig.~\ref{fig:hero_image}-C) enables the Core Agent to adapt its role to the active stage—guiding users, managing tasks, maintaining consistency, and reducing project-management overhead.
A stage-specific interface (Fig.~\ref{fig:hero_image}-B) divides the workflow into four dedicated boards—ideation, scripting, design, and storyboard—each tailored to its stage and functioning as shared memory for users and agents. A block- and element-based interaction model lets users select elements from outputs as references for new directions or revisions, supporting both divergent exploration and convergent refinement. Together, these features form an end-to-end workflow where creators collaborate with multiple agents to move seamlessly from initial concept to complete storyboard.

We conducted a summative study with 16 professional animation creators to evaluate AnimAgents. In a controlled within-subject experiment, using a strong single-agent baseline with parallel image generation (more powerful than participants’ typical tools), AnimAgents was rated significantly higher in coordination, consistency, information management, traceability, and overall satisfaction (all p < .01). In an open-ended task with participants’ own projects, AnimAgents was again preferred on those aspects. Participants valued AnimAgents as a project manager that coordinated tasks across stages, maintained visual and narrative consistency, organized outputs into clear, traceable boards, and reduced overhead, helping them stay focused on creative decisions and feel more in control.
A week-long field study with 4 professional creators showed that the value of AnimAgents diverges by context: increasing efficiency in commercial studios, facing limitations in original works, and scaffolding freelancers’ pre-production.

In summary, the major contributions of this work are:
\begin{itemize}
\item A formative study of end-to-end animation pre-production workflows with professional creative directors and independent animators, identifying the needs and challenges of coordinating across stages.
\item The design and implementation of AnimAgents, a stage-aware multi-agent orchestration system that mirrors studio role structures to support human–AI coordination across interdependent stages. The system introduces stage-specific boards, block- and element-based interactions, and information organization features that maintain continuity and enable traceable workflows across the entire pre-production pipeline.
\item A comprehensive multistage evaluation, including a controlled summative study and real-world field deployments, demonstrating AnimAgents' effectiveness in improving efficiency, consistency, information management, and creative agency.
\end{itemize}

\section{RELATED WORK}
\subsection{Agency in Human–AI Collaborative Creative Workflows} 

Generative AI (GenAI) is increasingly adopted in professional creative domains~\cite{zhou2024generative, amankwah2024impending}, with commercial systems now offering storyboard\footnote{\url{https://storyboardhero.ai/}
, \url{https://storyboarder.ai/}
, \url{https://boords.com/}}
 and video generation\footnote{\url{https://openai.com/sora/}
, \url{https://runwayml.com/}
, \url{https://www.midjourney.com/}}
for animation and film production. Yet these tools often prioritize efficiency and automation over human-in-the-loop control~\cite{izani2024impact, bourgault2023exploring}, risking diminished user agency~\cite{heer2019agency, epstein2023art}. In contrast, growing research investigates how GenAI can enhance creativity while preserving authorship and decision-making~\cite{eapen2023generative, watkins2024ai}. Recent HCI work advances human-centered co-creativity approaches that balance assistance with autonomy~\cite{moruzzi2024user, rezwana2025human, xu2023transitioning, auernhammer2020human}, ensuring creators remain active decision makers rather than passive recipients of AI output~\cite{moruzzi2024customizing, young2025balancing}.

Prior systems support creative workflows through divergent exploration in visual design~\cite{choi2025expandora, almeda2024prompting, choi2024creativeconnect} and storytelling~\cite{shaer2024ai, radwan2024sard, fu2025like}, or through convergent refinement to narrow options~\cite{wang2023popblends, PromptCharm2024, lin2025inkspire}. Others enable fine-grained control in image generation via interactive prompt editing~\cite{brade2023promptify, feng2023promptmagician, wang2023reprompt}, multimodal inputs~\cite{peng2024designprompt, lin2025sketchflex}, and interpretability features~\cite{chung2023promptpaint, mishra2025promptaid}. AIdeation~\cite{wang2025aideation} extends this by decomposing brainstorming outputs for transparency. Some works frame GenAI as a mediator for stakeholder communication~\cite{lobbers2023ai, chung2023artinter}, while others position users as curators of outputs~\cite{DesigningChiou2023}. Collectively, these strategies aim to preserve agency in creative tasks. Yet emerging workflows increasingly rely on automated pipelines, such as LLM chaining or agentic workflow~\cite{hong2024metagpt, wei2022chain}, which often limit users’ ability to understand or steer the process, especially in creative domains~\cite{wang2025gentune, wang2024lave}. Although effective for single-stage tasks, prior methods do not meet these challenges.

To address these limitations, recent research has helped users refine AI-generated intermediate outputs. DesignWeaver~\cite{tao2025designweaver} and Brickify~\cite{shi2025brickify} decompose visual elements into recombinable tokens to express the design intention.
GenTune~\cite{wang2025gentune} establishes explicit mappings between image elements and AI-generated prompts, providing traceable, element-level control.
For moving images, Keyframer~\cite{tseng2024keyframer} generates editable CSS animation code as an intermediate artifact, while MemoVis~\cite{chen2024memovis} turns textual feedback into AI-image references for refinement. Together, these tools provide finer-grained control over complex generative pipelines.

While these tools enhance control within specific tasks, they remain fragmented and lack an integrated framework for coordinating multi-stage generation in creative domains. AnimAgents addresses this gap by orchestrating control across stages through a central agent, while offering block- and element-level interactions that enable fine-grained control throughout the workflow.

\subsection{LLM-based Agentic Systems in Human Workflows}
While many LLM-based tools have been developed to support human workflows, prompting alone is insufficient for complex, multi-step processes~\cite{li2024autoflow, hong2024metagpt}. Recent work has thus introduced agentic systems that extend LLMs with memory, planning, and action capabilities for more autonomous execution~\cite{xi2025rise, wang2024llmagent}. For instance, single-agent systems have been applied to domain-specific automation, from software engineering~\cite{zhang2024codeagent, yang2024sweagent} to creative tasks~\cite{li2024anim}.

However, single-agent approaches remain limited for complex pipelines that demand diverse expertise~\cite{guo2024large, shen2024data, atmakuru2024cs4}. To address this, recent work has turned to multi-agent systems (MAS), where specialized agents coordinate through cooperative, competitive, or mixed strategies~\cite{tran2025multi, li2024massurvey} to solve tasks beyond a single agent’s scope~\cite{tao2024magis, su2025manyheads, ghafarollahi2024protagents}. For example, MapCoder~\cite{islam2024mapcoder} shows how specialized agents can tackle complex programming problems, while Agent Laboratory~\cite{schmidgall2025agent} demonstrates agents collaborating on research tasks. MAS have also entered creative domains: FilmAgent~\cite{xu2024filmagent} models agents for film production, AniMaker~\cite{shi2025animaker} emulates animation workflows for automated storytelling, and ReelWave~\cite{wang2025reelwave} coordinates agents for automated film sound. Yet despite their automation strengths, these systems often overlook human-in-the-loop collaboration, limiting creative control and refinement.

To address these gaps, recent HCI work emphasizes human–agent collaboration, designing systems that support rather than replace human workflows. Some employ a single agent working directly with users~\cite{peng2025morae, liu2024coquest}. For example, LAVE~\cite{wang2024lave} offers a collaborative video editing tool, while Trailblazer~\cite{yan2025answering} explores codebases to generate annotated traces for navigation. Others involve coordinating with multiple agents: ContextCam~\cite{fan2024contextcam} enables image co-creation through stage-specific agents, and DreamGarden~\cite{earle2025dream} assists game design by turning prompts into hierarchical plans with assets, code, and scenes in Unreal Engine.
Researchers have also begun exploring human–multi-agent collaboration in creative domains. Cinema Multiverse Lounge~\cite{ryu2025cinema} introduces persona-based agents to enrich film appreciation, while MapStory~\cite{gunturu2025mapstory} applies a dual-agent architecture to transform scripts into map-centric animations. PosterMate~\cite{shin2025postermate} simulates target audience feedback for poster refinement, and StorySage~\cite{talaei2025storysage} supports autobiographical writing through a multi-agent framework. Collectively, these works illustrate the potential of human–multi-agent co-creation.

Despite these advances, little research addresses domain-specific human–multi-agent collaborative systems for complex, multi-stage professional workflows. AnimAgents advances this space with a stage-aware orchestration framework, where a core agent coordinates specialized agents across pre-production stages. Integrated into end-to-end animation workflows, AnimAgents supports collaboration while ensuring consistency, traceability, and continuity throughout the pipeline.

\subsection{Managing Complexity in Multi-Stage Workflows} 
Creative pipelines such as animation pre-production generate large volumes of heterogeneous output, leading to information overload, difficulty tracing decisions, and loss of coherence across stages~\cite{li2011novel, liu2010preliminary, levy2010animation, frich2019mapping}. Prior HCI research has sought to reduce such complexity through interface design, visualization, and version management~\cite{frich2019mapping, remy2020evaluating}. Visual canvases help structure and compare ideas~\cite{martin2014pragmatic}, while version control systems trace revisions, reduce cognitive load, and support decision making~\cite{sterman2022towards}. Yet these methods address isolated tasks rather than ensuring coherence across interdependent workflow stages.

GenAI amplifies these challenges by generating vast text and image outputs in a single session, typically shown in linear chats where earlier results are quickly buried, making it hard to recover context or trace idea evolution~\cite{zhou2025examining, oppenlaender2023mapping}. To support more effective human–AI collaboration, recent HCI research has explored structured representations to better visualize and organize outputs~\cite{bodker1998understanding}. Many adopt graph-based interfaces~\cite{dang2023worldsmith, yan2023xcreation, wu2022promptchainer}. For example,  IdeationWeb \cite{shen2025ideationweb} tracks the evolution of design ideas using structured representations, analogical reasoning, and interactive visualization; Qin et al.\cite{qin2025toward} propose a node-graph interface for creative writing with customizable nodes and LLM audience impersonation; and SARD supports multi-chapter story generation via drag-and-drop narrative elements\cite{radwan2024sard}. Other systems embed domain knowledge into tailored interfaces, such as DesignPrompt~\cite{peng2024designprompt}, which extends prompting into visual modalities by combining sketches, diagrams, and structured inputs. For more complex multi-stage workflows, Cao et al.\cite{cao2025compositional} propose compositional structures as substrates to organize and visualize content for controlled creative processes. Collectively, these systems show how structured, domain-specific representations improve traceability and information management in GenAI-supported creativity.

Despite these advances, emerging GenAI systems increasingly pursue multi-stage agentic orchestration, where visualization alone is insufficient and coordination across stage outputs is critical for maintaining continuity and alignment. Recent tools explore this direction: some employ a single agent to manage multi-step outputs on visual canvases for image and video generation~\cite{lovart2025}, while others support non-linear co-design where human designers collaborate with AI agents across branching pathways~\cite{you2025designmanager}. ReelFramer~\cite{wang2024reelframer} similarly coordinates journalists and GenAI by using structured narrative and visual framings to guide content generation.

However, no existing tool helps creators manage overwhelming outputs or coordinate results across complex multi-stage workflows. AnimAgents addresses this gap with stage-specific interfaces featuring lineage and version tracking, coordinated by a project manager–like agent that oversees outputs across stages to enhance traceability, information management, and continuity.

\section{BACKGROUND: ANIMATION PRE-PRODUCTION WORKFLOW}
The animation production process is typically divided into three stages: 1) \textit{pre-production}, focused on story development, visual design, and planning; 2) \textit{production}, where assets are created and animated; and 3) \textit{post-production}, which finalizes the film through editing, sound, and music ~\cite{gossman2011animation, meroz2021animation, rall2017animation}.

\begin{figure}
    \centering
    \includegraphics[width=1\linewidth]{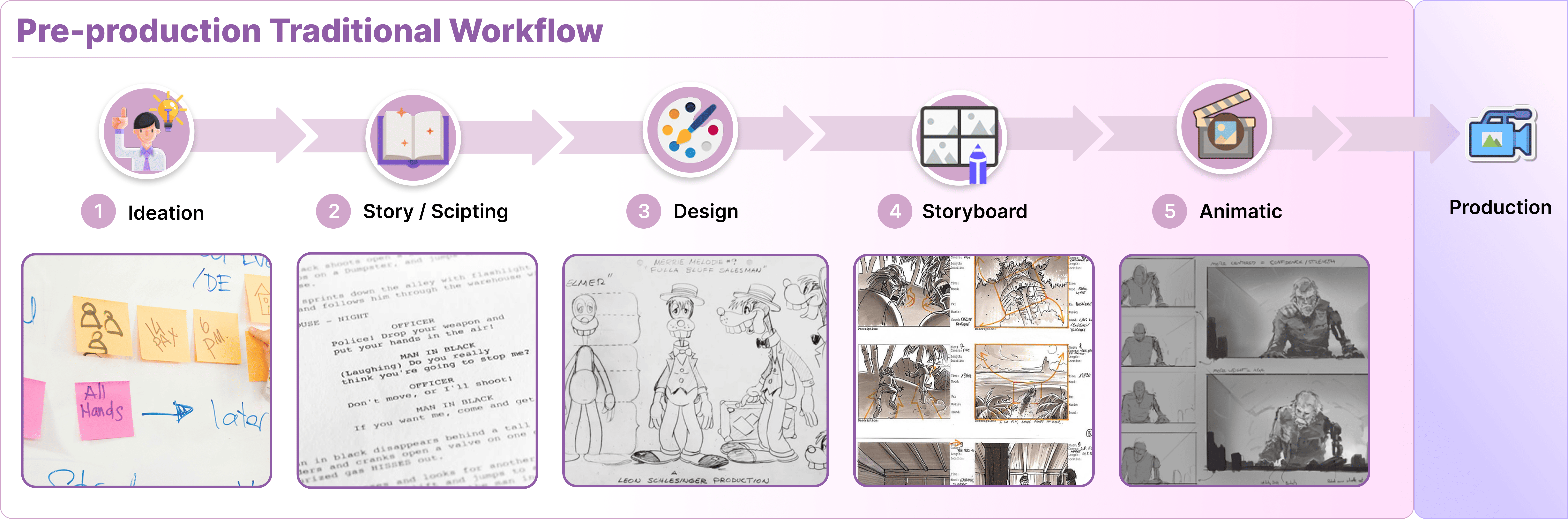}
    \caption{Typical traditional animation pre-production workflow, which progresses through five stages: Ideation, Story/Scripting, Design, Storyboard, and Animatic, before moving into production. The workflow is not strictly linear, as iterations and revisions may occur across stages.}
    \label{fig:preproduction_workflow}
    \Description[This figure shows a typical five-stage traditional animation pre-production workflow: Ideation, Story/Scripting, Design, Storyboard, and Animatic, leading to production.]{This figure shows the typical traditional animation pre-production workflow progressing through five stages before entering production stage. Each stage has a corresponding sample visual below it: Ideation – Sticky notes on a whiteboard, including sketches and written notes. Story / Scripting – A typewritten movie script page with dialogue and stage direction. Design – A hand-drawn character sheet showing character poses and outfits. Storyboard – pencil storyboard panels arranged in sequence, with perspective scenes and directional arrows. Animatic – Grayscale video stills showing characters and basic scene timing. Arrows connect each stage, forming a left-to-right horizontal flow toward the production stage. The caption notes that the process is not strictly linear, allowing for iteration and revision between stages.}
\end{figure}

Pre-production is the heart of an animated film~\cite{caldwell2001preproduction} (Fig.\ref{fig:preproduction_workflow}). It is organized into four iterative strands—\textit{ideation}, \textit{scripting}, \textit{design}, and \textit{storyboard}—that loop rather than proceed linearly. In \textit{ideation}, teams define the core vision (premise, genre, tone, theme, audience, constraints) and produce early artifacts such as a story concept and a “hero” keyframe for the look and feel. \textit{Scripting} shapes this concept into a structured narrative (e.g., setup–development–resolution), supported by a beat sheet, outline, and scene list, clarifying narrative style\cite{eich2024impact}. In \textit{design}, concept designers and art directors establish visual direction, characters, environments, and props, through model sheets, prop sheets, paintovers, and a concise art guide for downstream teams. \textit{Storyboard} turns the script into sequential shots specifying staging, composition, and timing; boards remain flexible and are frequently reordered~\cite{rall2017animation}. These strands converge in an \textit{animatic}—a timed rough cut integrating boards, temp audio, and provisional edits—to validate flow and pacing before full production~\cite{gwenaelle2022preproduction, kyle2023preproduction}.


In most studios, creative directors guide pre-production, collaborating with writers, storyboard artists, and designers to maintain creative authority while ensuring the evolving vision aligns with stakeholder expectations and production constraints~\cite{natasha2025what, anna2025what, noha2025role}.
In contrast, independent creators often assume multiple roles—concept development, scriptwriting, character and environment design, even budget planning. They produce diverse works such as socially driven shorts or web series~\cite{meegle2025animation}.
With tighter budgets, workflows are often streamlined. For example, storyboards may be simplified or skipped, as creators rely on mental visualization without the need for detailed team communication, moving directly into production~\cite{mike2025deepblueink}.

Although creative directors and independent animators approach pre-production differently, both follow an exploratory process from early ideation to concrete concepts. Increasingly, both also incorporate AI tools for tasks such as script analysis, automatic storyboard generation, and visual schematics, enabling faster iteration and shifting focus from manual groundwork to creative refinement~\cite{aiapps2025how, nazanin2025ai, clevertize2025how}.


\section{FORMATIVE STUDY}
We conducted a formative study to investigate current pre-production workflows in animation, examine how creators currently use GenAI tools, and identify the challenges they face as well as their expectations for future AI support.

\subsection{Participants}
We recruited 12 participants (1–16 YoE, Mean = 7.6): 6 creative directors from three animation studios (P1–6) and 6 independent animators (P7–12). Recruitment was through personal referrals in the local animation community and direct email to studios to request collaboration. Each participant received \$50 compensation. Table~\ref{tab:demographics} summarizes demographics, including role, experience, and prior GenAI exposure.

\subsection{Study Procedure}
Each participant completed a 1.5–2 hour semi-structured interview covering four topics: (1) pre-production workflow, (2) pain points and challenges, (3) current AI use and related challenges, and (4) expectations for AI assistance. To ground discussion, participants brought two past projects: one typical and one particularly challenging.

We prepared tailored question sets for creative directors and independent animators, with a shared core. Common questions addressed coordination and information management across pre-production. Director-specific questions emphasized cross-department communication, while animator-focused ones examined how they personally handled stage-specific challenges. Full interview protocols are in Appendix A.



\subsection{Findings}
We conducted a thematic analysis of the transcribed interviews. An author with prior professional experience in an animation studio developed the initial coding framework.
The coding process and resulting themes were iteratively discussed among four co-authors to ensure consistency and consensus.

All participants described pre-production as the most critical stage where story, art style, and visual direction are established, consistent with prior literature~\cite{meroz2021animation, rall2017animation}. While creative directors and independent animators approached the process differently, they reported facing similar challenges. 

\subsubsection{Roles of GenAI in Pre-production}
All participants reported using GenAI tools. LLMs were employed to expand story concepts and draft scripts (P1, P4, P8-11), while text-to-image models supported visual references and moodboards (P2, P4-5, P8-12). Creative directors mainly used these tools in early ideation to test directions before collaboration: “\textit{I used ChatGPT to expand different possible storylines}” (P1); “\textit{By quickly generating concept design references, I could imagine the story faster once I had visuals}” (P5). Independent animators also relied on GenAI to compensate for skill gaps: “\textit{Since I’m not good at storyboard, I generate it with AI, then block it out in 3D}” (P9), and “\textit{I let AI handle the script so I can move into visuals faster}” (P11).

\subsubsection{Challenges in Idea Development}
Participants found it especially difficult to transform abstract ideas into coherent stories (P2–4, P6–10, P12). As one noted, “\textit{At this stage, coming up with ideas is the hardest part, especially figuring out how to fill in and complete the story details}” (P6). Others struggled to write outside their own perspective (P7–8, P10). For instance, one reflected on portraying dementia: “\textit{When the script isn’t based on my own experience, it’s really hard to find the right angle}” (P7).

To address these difficulties, many participants turned to LLMs (P1, P3, P6–12), but the large volume of outputs often created more burden than benefit. Results were frequently trivial, requiring extensive filtering (P2–3, P6–8, P11): “\textit{Most of what the AI generates is pretty mediocre, and I have to spend a lot of time filtering}” (P11). Others noted irrelevant content, “\textit{AI often goes off on its own and produces a bunch I never asked for}” (P7), and the difficulty of organizing overwhelming information: “\textit{When brainstorming with AI, there’s so much output it’s really hard to sort through}” (P3).



\subsubsection{Challenges with Continuity Across Stages}
Another major challenge was maintaining continuity as projects moved through story, design, and visual development. Participants struggled to manage outputs and track progress (P1–3, P5–8, P10–12): “\textit{There’s just too much material across stages, and it’s hard to keep track of progress and remember earlier decisions}” (P3), and “\textit{I often organize everything on one board, but once there’s too much, it quickly gets messy}” (P9). Consistency in narrative voice, characters, and visual style also broke down over time (P1, P3–5, P8–11): “\textit{The further the project goes, the harder it is to stay consistent—it’s easy to forget the earlier narrative style}” (P8).

GenAI further amplified inconsistency, with outputs drifting in story or style: “\textit{When I keep working with AI, the new content often drifts away from the original story}” (P6). Managing variations added cognitive load and made it harder to trace earlier ideas (P2–4, P6, P10, P12): “\textit{It’s difficult to keep the narrative consistent when so many variations are generated}” (P5). While GenAI improved efficiency for single tasks, fragmentation across tools (scripting, design, image generation) undermined coherence: “\textit{Even when I asked ChatGPT to generate prompts in the same style, the images from MidJourney still didn’t match}” (P12).

Tracing earlier outputs was difficult, as AI tools lacked effective version tracking, making it easy to lose the original intent (P1, P5–6, P8, P10–11). “\textit{After discussing with ChatGPT for a while, it’s hard to find what I said earlier, I have to keep scrolling up}” (P10). Once substantial outputs accumulated across stages, refining earlier intermediate results became harder: “\textit{When I try to revise something from the early stage after progressing a lot, the new outputs don’t match, and I often have to redo everything from scratch}” (P8).

\subsubsection{Communication Burden}
Participants also struggled to share outputs in ways others could easily grasp. For creative directors, this meant communication overload when coordinating teams under time pressure (P2–4, P6): “\textit{It’s extremely difficult to sort and organize all the information. How do you communicate it clearly in meetings and share so much in limited time?}” (P5). Independent animators faced similar issues when presenting evolving ideas to clients (P7–8, P11): “\textit{It’s hard to present progress clearly, especially early on when clients change their minds and jump between versions}” (P7).



\subsubsection{Expectations for GenAI Assistants}



Across both groups, participants saw GenAI less as automation and more as a partner in coordinated creativity. They expected AI to follow instructions (P2–3, P7–8), expand ideas (P5, P8, P11), refine outputs (P3–4, P11), and offer suggestions (P2–3, P5, P8), while keeping direction in the creator’s hands (P5, P7–9, P11). Some, however, preferred to exclude AI from ideation entirely (P6).

More importantly, participants emphasized the need for project management support (P3, P6–7, P11). “\textit{What we lack most isn’t more creative ideas. It’s planning, organizing information, and telling me what’s still missing}” (P7). They also stressed coordinating results across stages (P2–3, P5–8, P10): “\textit{The most troublesome part is integrating all the information from every creative discussion}” (P3). Version control and structured visualization were frequently requested, including clearer ways to organize iterations—“\textit{If AI could clearly organize each version, it would be a huge help}” (P5), and artist-friendly interfaces: “\textit{Board-based or panel-style interfaces are more intuitive than text logs}” (P9). The creative directors also highlighted the reduction of the communication burden by automatically sharing updates across departments: “\textit{If the AI could distribute updates automatically, it would save many meetings}” (P3).



In summary, our formative study shows that creators struggle with ideation as well as coordinating and maintaining consistency across pre-production. 
Although current GenAI tools often add to these burdens, participants envisioned future systems as collaborative assistants supporting coordination and structured visualization. These findings highlight the need for a stage-aware system that manages information flow and sustains continuity across the pre-production pipeline.


\subsection{Design Goals}
Based on the findings, we proposed three design goals for our system:
\begin{itemize}
    \item \textbf{DG1: Ensure continuity and traceability across multi-stage workflows} 
    The system should help creators organize outputs, track progress, and revisit earlier decisions, while maintaining narrative and visual consistency across stages and providing clear traceability of prior versions and creative intentions (Section 4.3.3, 4.3.5).

    \item \textbf{DG2: Provide structured visualizations that align with creative practice}
    The system should present stage-specific visualizations that mirror artistic practice, enabling creators to navigate, compare, and trace outputs for effective decision making while also reducing communication burden when sharing with teams or clients (Sections 4.3.3-4.3.5).
    
    \item \textbf{DG3: Expand creative possibility while preserving agency} 
    The system should support exploration and refinement while minimizing trivial tasks (e.g., filtering mediocre outputs, project management). By faithfully following user instructions and letting creators choose the level of AI involvement, it ensures key creative decisions remain human-led, allowing them to focus on core creative tasks (Sections 4.3.1-4.3.2, 4.3.5).
    
    

\end{itemize}

\begin{table*}[h!]
\centering
\small
\begin{tabular}{|c|c|c|c|c|c|c|c|c|}
\hline
\textbf{ID} & \textbf{Age} & \textbf{YoE} & \textbf{Industry} & \textbf{Role} & \textbf{GenAI Tools Used} & \textbf{Formative} & \textbf{Summative} & \textbf{Field} \\ \hline
1 & 40 & 16 & Animation & Creative Director & ChatGPT & \checkmark & & \\ \hline
2 & 38 & 12 & Animation & Creative Director & ChatGPT & \checkmark & & \\ \hline
3 & 38 & 15 & Animation & Creative Director & ChatGPT & \checkmark & & \\ \hline
4 & 39 & 15 & Animation & Creative Director & ChatGPT & \checkmark & & \\ \hline
5 & 37 & 12 & Animation & Creative Director & ChatGPT & \checkmark & & \checkmark \\ \hline
6 & 29 & 7 & Animation, Freelancer & Creative Director & ChatGPT & \checkmark & & \\ \hline
7 & 23 & 1 & Graduate Student, Freelancer & Independent Animator & ChatGPT & \checkmark & & \\ \hline
8 & 26 & 2 & Animation, Freelancer & Independent Animator & Midjourney, ChatGPT & \checkmark & \checkmark & \checkmark\\ \hline
9 & 25 & 2 & Animation, Freelancer & Independent Animator & ChatGPT & \checkmark & \checkmark & \\ \hline
10 & 23 & 1 & Graduate Student, Freelancer & Independent Animator & ChatGPT & \checkmark & \checkmark & \\ \hline
11 & 27 & 5 & Animation, Freelancer & Independent Animator & ChatGPT & \checkmark & \checkmark & \\ \hline
12 & 27 & 3 & Animation, Freelancer & Independent Animator & ChatGPT, Flux & \checkmark & \checkmark & \\ \hline
13 & 24 & 3 & Animation, Freelancer & Creative Director & ChatGPT & & \checkmark & \\ \hline
14 & 24 & 1 & Animation, Freelancer & Creative Director & Stable Diffusion, ChatGPT & & \checkmark & \\ \hline
15 & 39 & 13 & Animation & Creative Director & Midjourney, ChatGPT & & \checkmark & \\ \hline
16 & 25 & 4 & Animation & Creative Director & Midjourney, ChatGPT & & \checkmark & \\ \hline
17 & 31 & 7 & Animation & Independent Animator & Midjourney, ChatGPT, Runway & & \checkmark & \\ \hline
18 & 31 & 7 & Animation, Freelancer & Independent Animator & ChatGPT, Grok & & \checkmark & \\ \hline
19 & 32 & 8 & Animation, Freelancer & Independent Animator & ChatGPT & & \checkmark & \checkmark \\ \hline
20 & 33 & 7 & Animation, Freelancer & Independent Animator & Midjourney, ChatGPT, Leonardo & & \checkmark & \\ \hline
21 & 26 & 4 & Animation, Freelancer & Independent Animator & Midjourney, ChatGPT, Krea & & \checkmark & \\ \hline
22 & 27 & 4 & Animation, Freelancer & Independent Animator & Stable Diffusion, ChatGPT & & \checkmark & \\ \hline
23 & 25 & 3 & Graduate Student, Freelancer & Independent Animator & Stable Diffusion, ChatGPT, Flux & & \checkmark & \\ \hline
24 & 36 & 12 & Animation & Creative Director & ChatGPT & & & \checkmark \\ \hline
\end{tabular}
\caption{Demographic Details of Participants Including Age, Years of Experience, Industry, Role, Generative AI Tools Usage, and Study Participation.}
\label{tab:demographics}
\Description[This table shows demographic details of 24 participants, including age, years of experience, industries, roles, generative AI tools used, and study participation.]
{This table presents demographic details of 24 participants in the study, highlighting their backgrounds, tool usage, and study involvement. Participants ranged in age from 23 to 40 years old, with professional experience spanning 1 to 16 years. Most worked in the animation industry, and their roles were primarily divided between Creative Directors and Independent Animators. Creative Directors tended to have more than 10 years of experience, while Independent Animators were often early-career professionals or freelancers with 1 to 7 years of experience. A few participants were graduate students combining their studies with freelance animation work.
In terms of generative AI tools, every participant used ChatGPT, either alone or in combination with other tools such as Midjourney, Stable Diffusion, Flux, Runway, Leonardo, Krea, or Grok. Patterns of usage also differed: Creative Directors often relied on ChatGPT alone or alongside Midjourney, whereas Independent Animators experimented with a wider range of tools.
The table also shows participation across study phases. Many Creative Directors were involved in the Formative study, while Independent Animators contributed more heavily to the Summative and Field studies. A few individuals, such as P8, P19, and P24, participated across multiple phases.
Overall, the demographics reflect a diverse participant pool that balanced senior professionals with early-career animators, all of whom engaged with generative AI tools in varied ways.}
\end{table*}

\section{SYSTEM \& IMPLEMENTATION}

We present AnimAgents, a human–multi-agent collaborative system for end-to-end animation pre-production. It supports both the exploratory workflows of studio directors and the hands-on practices of independent creators, enabling iterative loops from ideation to scripting, design, and storyboard with cross-stage continuity. Unlike prior single-agent or multi-window workflows, AnimAgents introduces stage-aware orchestration, where a Core Agent coordinates Specialized Agents and organizes results on stage-specific boards.

\subsection{System Overview}
AnimAgents models the workflow of a creative director, with agents mirroring cross-department collaboration in an animation studio. Responsibilities are distributed across a Core Agent, acting as project manager, and four Specialized Agents (Ideation, Scripting, Design, Art). This division keeps each agent’s memory focused, reducing confusion and preserving cross-stage consistency. The workflow spans five stages: Planning, Ideation, Scripting, Design, and Storyboard, with the Core Agent adapting to the active stage, tailoring interactions, and allowing flexible stage switching. Planning is handled solely by the Core Agent, which records project information. Each subsequent stage has its own board, where outputs are stored as lineage-aware blocks that capture multiple versions and trace their evolution (Fig. \ref{fig:hero_image}-B).

\begin{figure*}[htbp]
    \centering
    \includegraphics[width=0.98\linewidth]{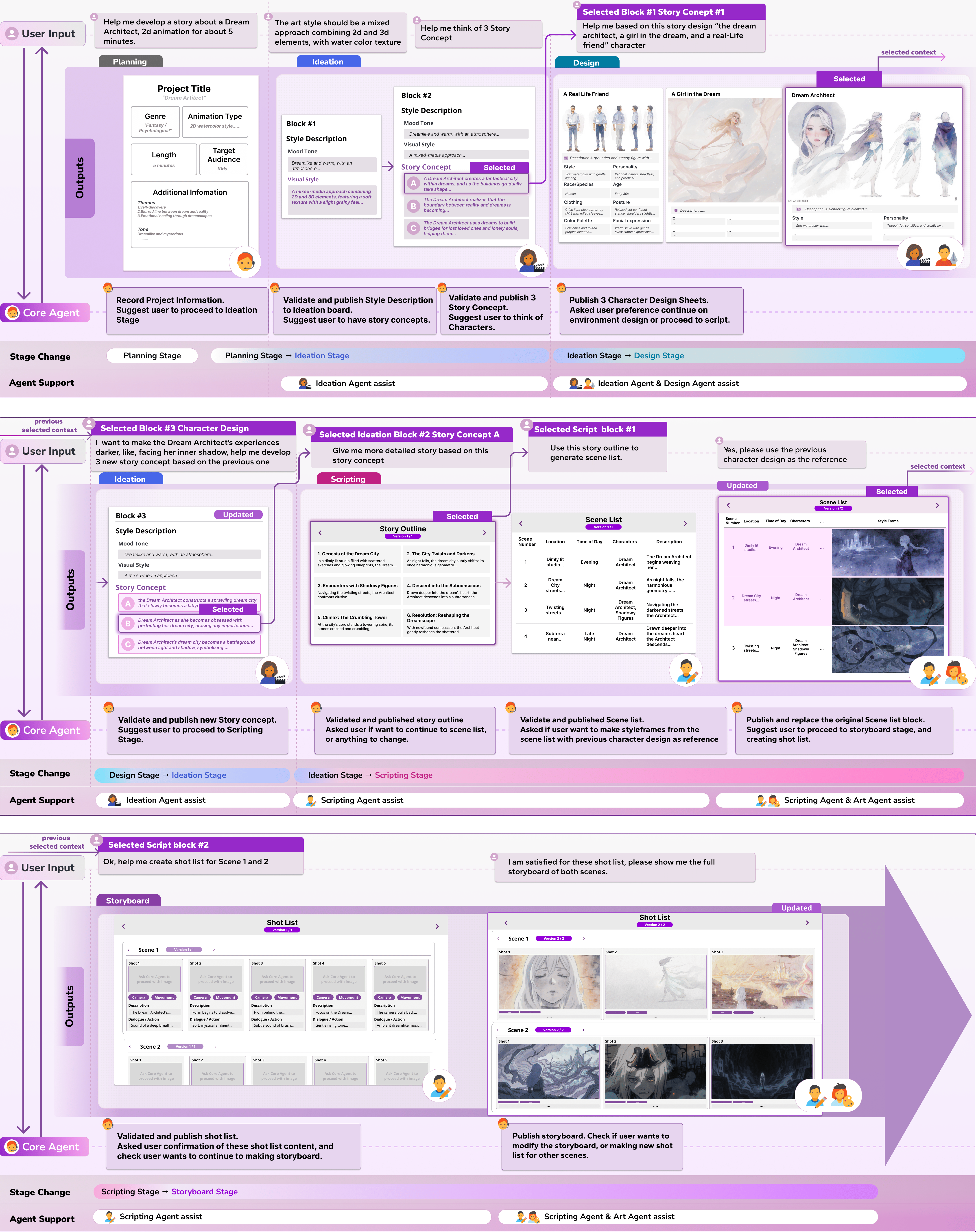}
    \caption{AnimAgents' workflow from P22 in the summative study. The user began by initiating a conversation with the Core Agent, describing a story idea about “Dream Architect.” The workflow progressed through the stages of Planning, Ideation, Design, a second round of Ideation, Scripting, and Storyboard, ultimately resulting in a completed shot list and storyboard.}
    \label{fig:system_workflow}
    \Description[This figure shows the AnimAgents workflow from participant P22, progressing from a story idea to a completed shot list and storyboard.]{This figure shows the workflow of participant P22 using AnimAgents in the summative study. The process began with a conversation with the Core Agent, where the user proposed a story idea about “Dream Architect.” From there, the workflow advanced sequentially through multiple stages. Planning was followed by Ideation, in which creative concepts were generated. The process then moved to Design, where visual elements were developed, before returning to a second round of Ideation for refinement. After this, the Scripting stage produced narrative structures and detailed scenes. Finally, the Storyboard stage generated a shot list and complete storyboard. The figure illustrates how AnimAgents supports iterative development across stages, culminating in fully structured narrative outputs.}
\end{figure*}

We illustrate the workflow with an example from our summative study (Fig.~\ref{fig:system_workflow}). A creator began in the Planning stage with a project brief: “a 5-minute 2D animation about a dream architect.” The Core Agent recorded it, proposed next steps, and initiated Ideation. The user specified a visual style, and requested three story concepts. The Core Agent created task specifications and delegated this task to the Ideation Agent, which documented the corresponding style directions and expanded the brief into three alternatives, all stored as blocks on the Ideation board. 

AnimAgents supports element-level selection, letting users refine or extend specific outputs. In the example, the creator chose one story concept and specified three characters: “the dream architect, a girl in the dream, and the real-life friend.” The Ideation Agent generated character concepts, which, once approved, the Core Agent passed to the Design Agent with specifications and context from Ideation. The Design Agent then produced visual character designs.

After reviewing the designs, the Core Agent suggested moving to scripting, but the creator branched the story by making one of the Dream Architect’s experiences darker, adding a new branch on the Ideation board. This illustrates that agents can be re-engaged iteratively rather than in sequence, and users can switch stages flexibly. The Core Agent integrated the change and advanced to Scripting, passing prior designs and the revised concept as context. The Scripting Agent produced an outline and scene list. The user then selected a scene and requested a storyboard. In Storyboard, the Scripting Agent mapped the narrative into shots, while the Art Agent used prior designs and the user’s style description to generate aligned outputs.

The Core Agent serves as the central coordinator: interpreting instructions, delegating tasks, tracking progress, and suggesting next steps. It adapts to each stage: encouraging divergent exploration in Ideation, enforcing continuity in Scripting, and checking visual–narrative consistency in Storyboard. This stage-based orchestration prevents overload and ensures precise, context-aware execution. Combined with stage-specific boards and element-level refinement, AnimAgents supports iterative generation, review, and revision while preserving coherence from concept to storyboard.


\begin{figure*}[htbp]
    \centering
    \includegraphics[width=1\linewidth]{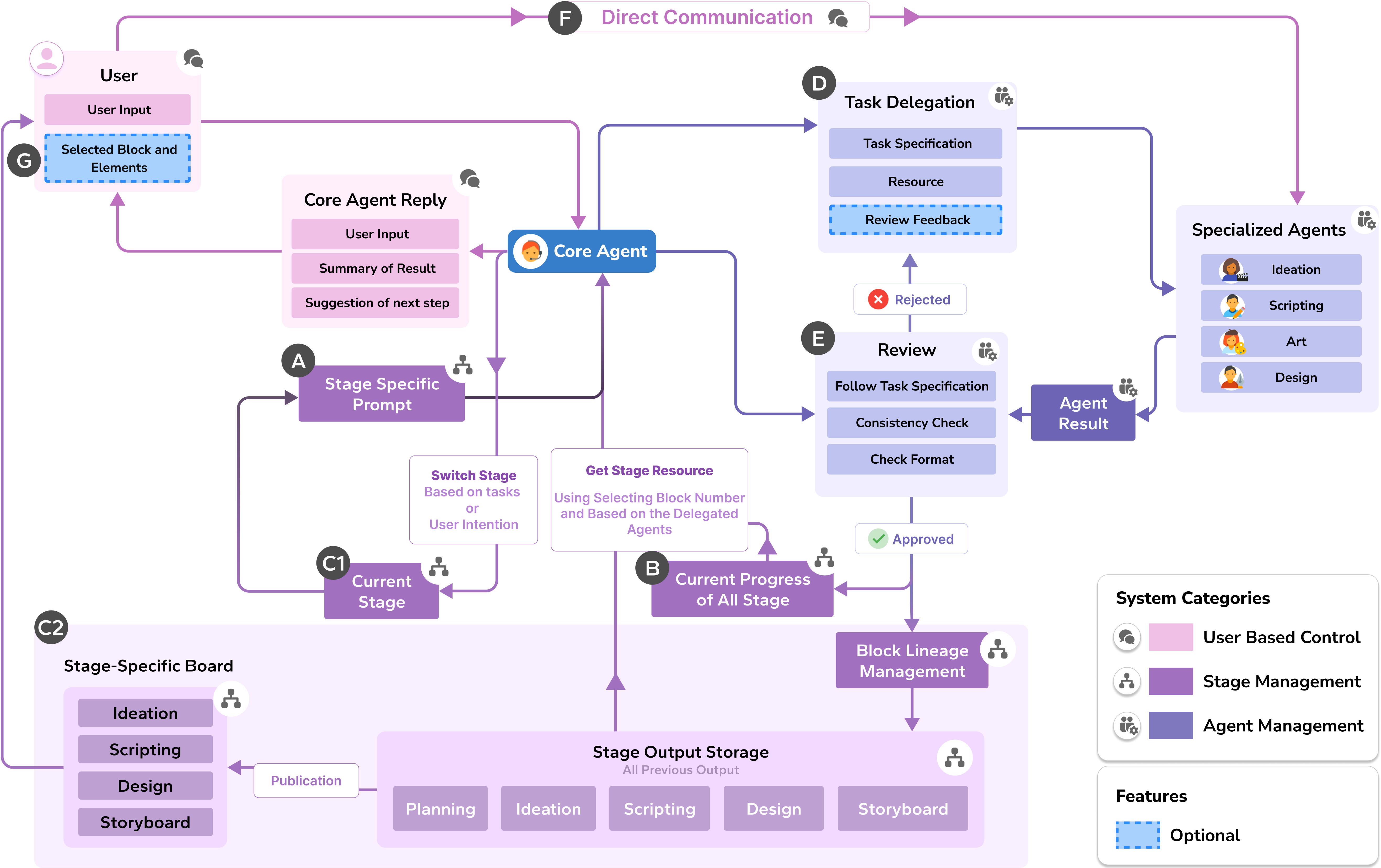}
    \caption{System architecture of AnimAgents. For stage management, the Core Agent determines the (C1) current stage based on user input, provides (A) stage-specific prompts, and (B) tracks progress based on agent outputs. Results are then displayed on (C2) stage-specific boards. For agent management, it (D) delegates tasks to Specialized Agents and (E) reviews their results before approval. Users interact with agents through text and (G) selected blocks and elements, and can (F) directly communicate with Specialized Agents.}
    \label{fig:multi-agent-architecture}
    \Description{This figure shows the system architecture of AnimAgents, a human–multi-agent orchestration framework for animation pre-production. The workflow is illustrated as a flow diagram with labeled components and directional arrows, grouped into three categories: user-based control, stage management, and agent management.  On the left, the User provides input and can select blocks or elements (G). The Core Agent at the center manages stage progression. It issues (A) stage-specific prompts, identifies (C1) the current stage, and tracks (B) current progress. Results are displayed on (C2) stage-specific boards for Ideation, Scripting, Design, and Storyboard, connected to stage output storage for all previous outputs.  For Agent Management, the Core Agent (D) delegates tasks with specifications and resources to Specialized Agents (Ideation, Scripting, Art, Design). These agents return results, which then go through (E) review: task specification check, consistency check, and format validation. Rejected results loop back for revision; approved results proceed to storage and lineage management.  Users can also (F) directly communicate with specialized agents, in addition to interacting via the Core Agent. Colored labels indicate categories: pink for user-based control, purple for stage management, blue-purple for agent management, and dashed blue for optional elements.}
\end{figure*}

\subsection{Stage-aware Multi-Agent Architecture}
AnimAgents adopts a cooperative, role-based, and centralized architecture ~\cite{tran2025multi}, with a stage-aware Core Agent orchestrating Specialized Agents (Fig.~\ref{fig:multi-agent-architecture}). This balances centralized coordination with distributed task execution, ensuring workflow consistency and parallel creativity.

Reflecting certain real-world studio practices, Specialized Agents do not communicate directly with one another; instead, the Core Agent distributes and collects all tasks and outputs, mirroring a project manager channeling information across departments.

\subsubsection{Core Agent}
The Core Agent functions as the project manager, guided by a base system prompt defining its high-level responsibilities, stage definitions, management rules, delegation logic, and user interaction guidelines. It also loads stage-specific prompts (Fig.~\ref{fig:multi-agent-architecture}-A) specifying responsibilities, tasks, delegation procedures, and validation checks to ensure alignment with the active stage.

Core Agent's responsibilities include:
\begin{itemize}
    \item \textbf{Progress Tracking.} Maintains a record of artifacts and progress at each stage, serving as a resource for delegation and to ensure consistency within and across stages (Fig. \ref{fig:multi-agent-architecture}-B). When a result is approved, it either replaces the existing artifact of the same type (e.g., a new story outline) or is added if no equivalent exists.
    \item \textbf{Stage management and coordination.} Suggests stage-specific tasks and next steps, seeking user approval before major actions. Actively switches stages when requested by the user or when outputs meet stage completion criteria defined in the project context (Fig.~\ref{fig:multi-agent-architecture}-C).
    \item \textbf{Task delegation.} Translates user requests and decides whether to respond directly or delegate to one or more special agents  (Fig. \ref{fig:multi-agent-architecture}-D). When delegating, it generates a task specification and packages relevant context and resources from earlier stages to maintain consistency. 
    \item \textbf{Result validation and publication.} The Core Agent reviews Specialized Agent outputs against the task specification and project context to check formatting, compliance, and cross-stage consistency (Fig.~\ref{fig:multi-agent-architecture}-E). If validation fails, it provides feedback and requests revision. Once approved, the result is published as a block on the relevant stage board, either as a child of a parent block or by overwriting an existing one, and summarized for the user in chat.
    \item \textbf{Interaction modes.} Users can work in selection mode, choosing a specific block as context (Fig.~\ref{fig:multi-agent-architecture}-G) and passing it to an agent, or in regular mode, where the Core Agent defaults to the most recent result.

\end{itemize}

\begin{table*}[t]
\centering

\begin{tabular}{|c|p{7cm}|p{8cm}|}
\hline
\textbf{Role} & \textbf{Description} & \textbf{Primary Tasks} \\
\hline
\textbf{Ideation Agent} & Divergent and generative; explores multiple narrative directions and expands creative possibilities. & Generate story ideas, character concepts, world settings, narrative themes, and style descriptions. \\
\hline
\textbf{Scripting Agent} & Convergent and structured; transforms abstract concepts into coherent narrative sequences. & Produce structure mapping, story outlines, scene lists, and full scripts; adapt scripts based on selected ideas and design context. \\
\hline
\textbf{Design Agent} & Visual and style-driven; transform concept into concrete visual design. & Create character design sheets and environment designs; maintain visual coherence with narrative intent. \\
\hline
\textbf{Art Agent} & Illustrative and compositional; translates scripts and designs into polished visual assets. & Generate hero images, styleframes, and storyboards; align visual composition with narrative and design requirements. \\
\hline
\end{tabular}
\caption{Responsibilities of Specialized Agents in AnimAgents.}
\label{tab:specialized_agents}
\Description[This table shows the responsibilities of four specialized agents in AnimAgents: Ideation, Scripting, Design, and Art.]
{This table describes the responsibilities of four specialized agents in AnimAgents.  
The Ideation Agent is divergent and generative. It explores multiple narrative directions and expands creative possibilities. Its primary tasks include generating story ideas, character concepts, world settings, narrative themes, and style descriptions.  
The Scripting Agent is convergent and structured. It transforms abstract concepts into coherent narrative sequences. Its tasks include producing structure mappings, story outlines, scene lists, and full scripts, and adapting scripts based on selected ideas and design context.  
The Design Agent is visual and style-driven. It transforms concepts into concrete visual design. Its main tasks are creating character design sheets and environment designs while maintaining visual coherence with the narrative intent.  
The Art Agent is illustrative and compositional. It translates scripts and designs into polished visual assets. Its tasks include generating hero images, styleframes, and storyboards, and ensuring that visual composition aligns with narrative and design requirements.}
\end{table*}

\subsubsection{Specialized Agent}
Each Specialized Agent runs under a role-specific system prompt and tailored tools with task-specific prompts. These define responsibilities, behavior, and tool usage. For example, the Scripting Agent includes a  \texttt{make\_scene\_list} tool that takes a task description, message context, and prior-stage input (e.g., a story outline) to generate structured outputs with an LLM. Under Core Agent mediation, Specialized Agents perform only assigned tasks, ensuring controlled, stage-appropriate behavior.

Their roles span ideation, scripting, design, and art, with responsibilities summarized in Table~\ref{tab:specialized_agents}. This division keeps agents focused and memory contexts manageable. Beyond specialization, agents support two features:
\begin{itemize}
    \item Parallel execution. Multiple agents can generate variations simultaneously and return them for Core Agent review.
    \item Direct communication. On request, the Core Agent opens a chat with a chosen agent by forwarding project context and switching to its chatroom (Fig.~\ref{fig:multi-agent-architecture}-F). The Specialized Agent can interact directly with the user, while all tool calls and outputs route back through the Core for validation, lineage, and publication. This mode is useful when users want to provide detailed context or when tasks are too complex for automatic delegation.
\end{itemize}


This architecture ensures that the Core and Specialized Agents only perform user requested tasks, reducing irrelevant outputs and manual filtering. The Core Agent enforces alignment with instructions, maintains cross-stage consistency, and guides workflow progress. The modular structure also allows new agents and tools to be added without altering the protocol.

\begin{figure*}[htbp]
    \centering
    \includegraphics[width=1\linewidth]{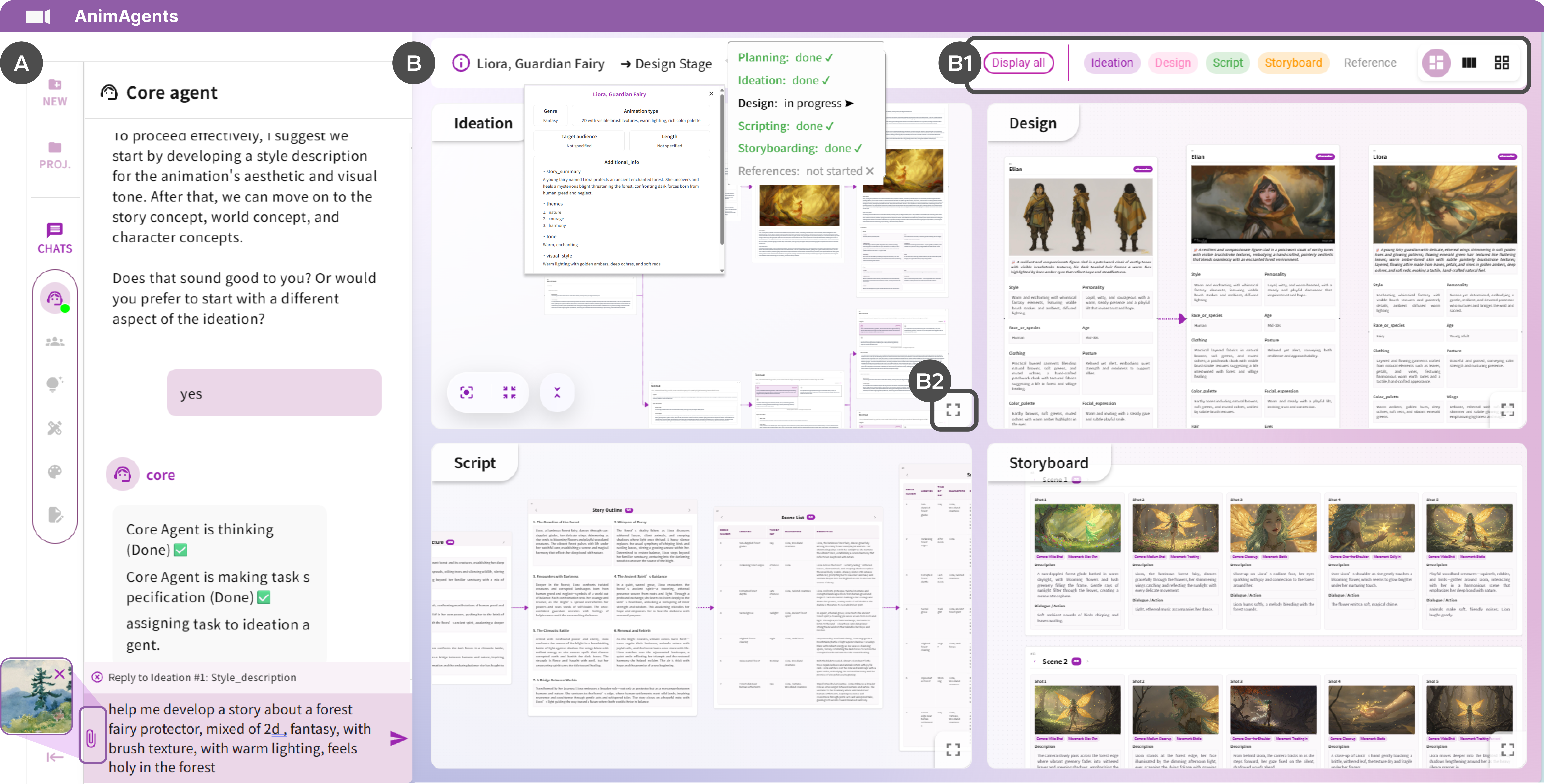}
    \caption{Main interface of AnimAgents. (A) Chat panel, where users can chat with agents, interact through conversation, and upload images. (B) Board workspace on the right, consisting of four infinite canvases (boards)—Ideation, Design, Scripting, and Storyboard—with a progress bar at the top showing the current stage progress and project information panel. Users can toggle which boards are displayed (B1) and expand or collapse an individual board (B2).}
    \label{fig:ui}
    \Description[This figure shows the main interface of AnimAgents, with a chat panel and four interactive boards for ideation, design, scripting, and storyboard stages.]{This figure shows the main interface of AnimAgents, a collaborative animation pre-production platform. The interface is divided into two main regions:  (A) On the left is the chat panel, where users interact with the Core Agent through conversation. Users can also upload reference images in the conversation.  (B) On the right is the board workspace, consisting of four horizontally aligned infinite-scroll boards: Ideation, Design, Scripting, and Storyboard. Each board contains structured content, such as scene outlines, character sheets, scripts, and visual storyboards.  At the top of the board area is project information and a stage progress tracker, showing the status for current stages.  In the upper-right section (B1), a row of color-coded toggle buttons allows users to show or hide individual boards. In the lower-right corner of each board (B2), a fullscreen button is provided to expand/collapse specific boards.}
\end{figure*}

\subsection{User Interface and Output Visualization}
AnimAgents provides a multi-board canvas interface that organizes outputs by pre-production stage (Fig.\ref{fig:ui}). By default, all boards are expanded for a full workflow view, but each can be opened or collapsed to focus on a single stage (Fig.\ref{fig:ui}-B1, B2).

\subsubsection{Stage-Specific Boards and Outputs}
Each board organizes results into lineage-aware blocks that capture revisions and branches, allowing creators to compare, refine, and continue concepts while preserving alternatives. The boards use a Figma-like infinite canvas for free navigation, with new blocks auto-positioned and stage-specific block structures applied. 

\begin{figure}[htbp]
    \centering
    \includegraphics[width=1\linewidth]
    {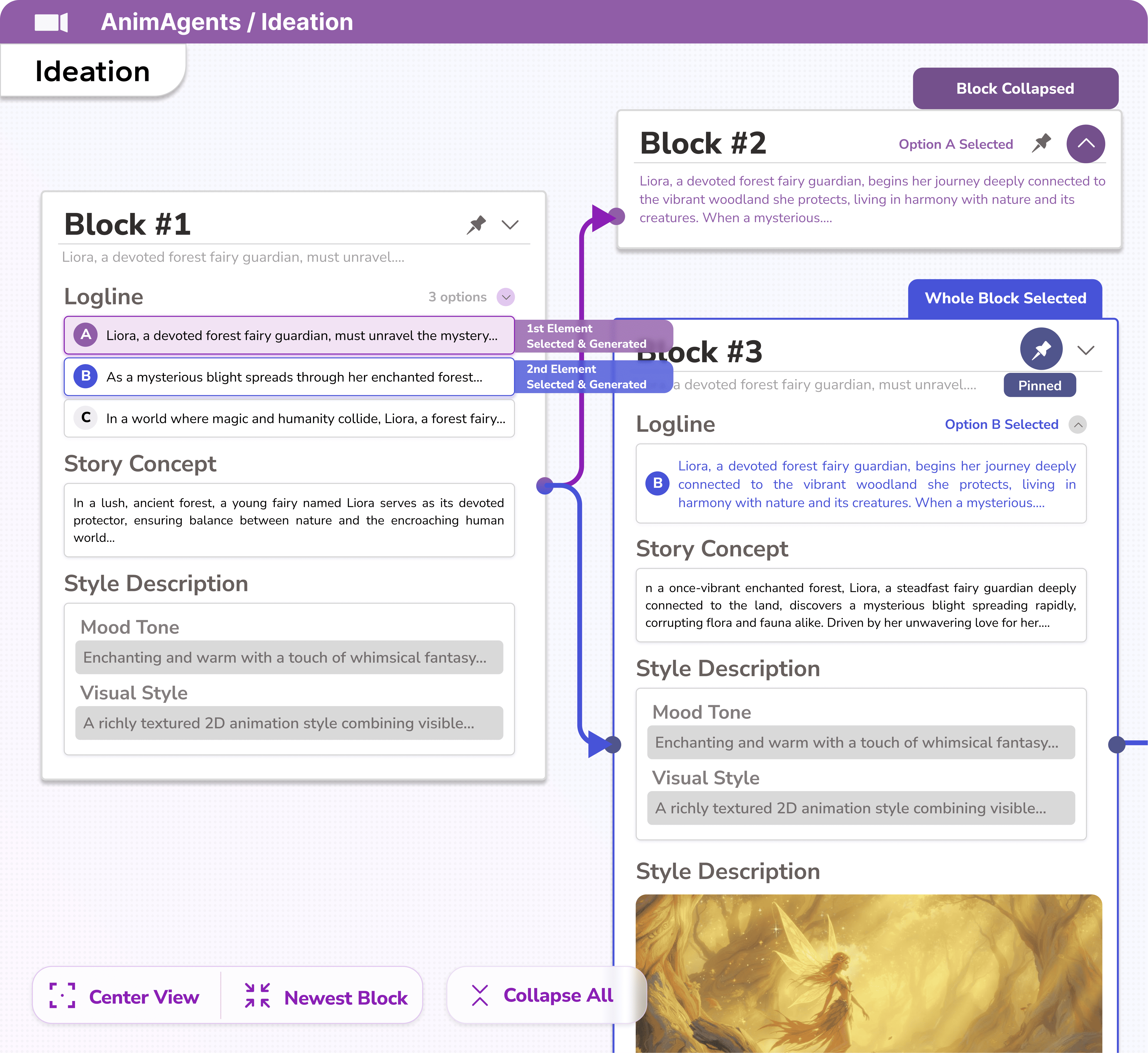}
    \caption{Ideation board in AnimAgents. Each block contains multiple options (e.g., loglines) for user selection and can be collapsed, pinned, or expanded. When a block is modified or extended, a new branch is created, preserving previous content and enabling flexible, iterative exploration.}
    \label{fig:ideation_board}
    \Description[This figure shows the ideation board of AnimAgents, where users can select individual elements within a block or choose the entire block; and expand, collapse, or pin each block for flexible exploration]{This figure shows the Ideation board in AnimAgents, a canvas for displaying and refining generated ideas. The board is organized into blocks, each containing structured creative elements such as loglines, story concepts, and style descriptions. Users can select individual elements within a block or select the block as a whole. Blocks can be expanded, collapsed, or pinned, and selections generate new branches while preserving previous content. This branching structure enables iterative and flexible exploration of creative directions. At the bottom of the interface, controls are provided to center the view, highlight the newest block, or collapse all blocks.}
\end{figure}

\begin{figure}[htbp]
    \centering
    \includegraphics[width=1\linewidth]
    {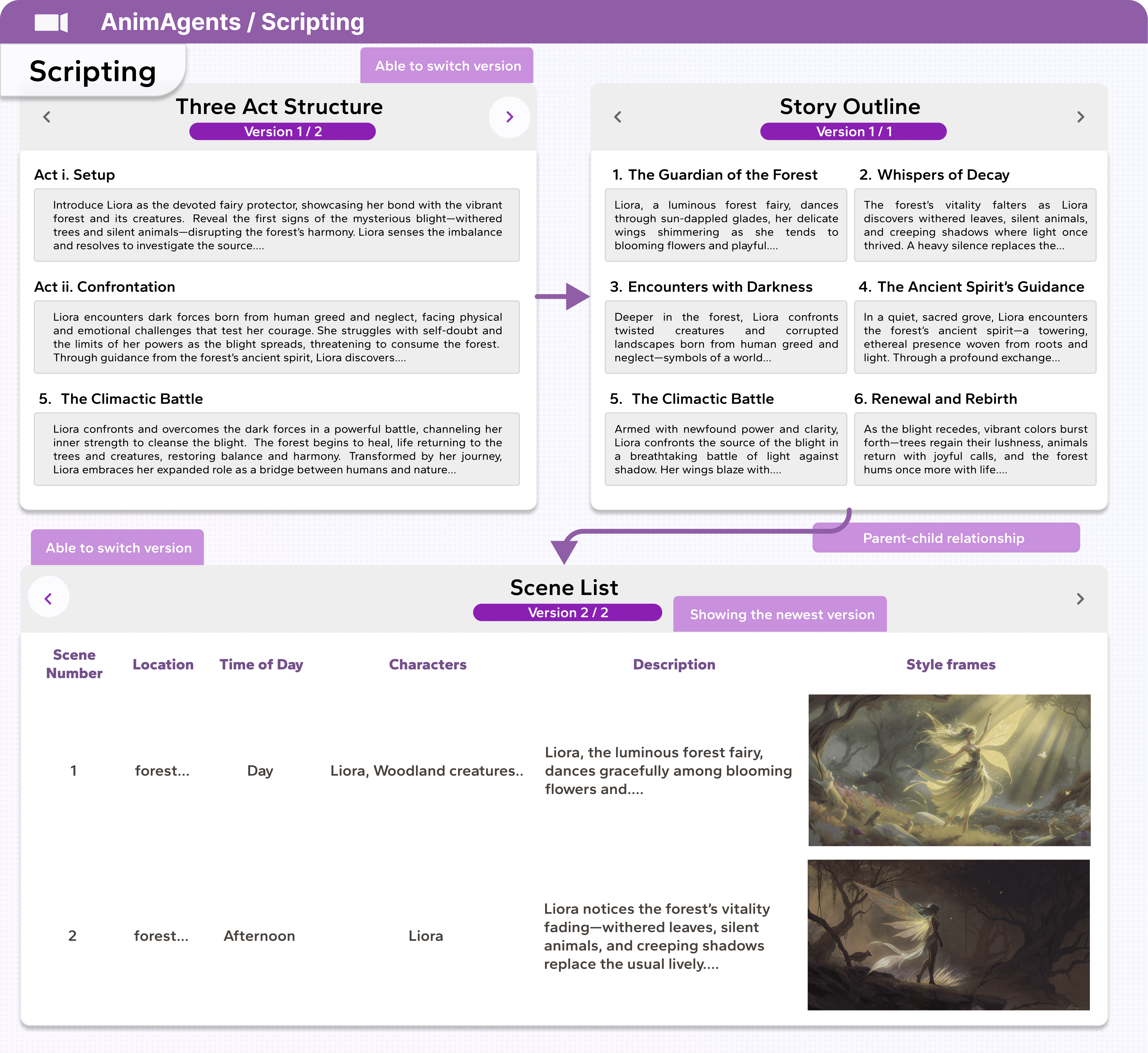}
    \caption{Scripting board in AnimAgents supports structured narrative development through three block types: Three Act Structure, Story Outline, and Scene List. Each refinement generates a new version; users can switch between versions, preserving prior drafts for comparison and traceability.}
    \label{fig:scripting_board}
    \Description[This figure shows the Scripting board in AnimAgents, where users develop narratives using Three Act Structure, Story Outline, and Scene List blocks with version control.] {This figure shows the Scripting board in AnimAgents, which supports structured narrative development through three types of blocks: Three Act Structure, dividing the story into Acts I, II, and III with key turning points; Story Outline, summarizing the sequence of story events; and Scene List, enumerating individual scenes with details of scene number, location, time of day, characters, description, and style-frame.
    Each refinement or modification produces a new version of the block while preserving previous drafts, allowing users to switch between versions for comparison and traceability. The interface highlights version history and enables iterative narrative construction.}
\end{figure}

The Ideation board emphasizes divergent exploration: the Ideation Agent generates multiple options per task, shown as parallel alternatives (Fig.\ref{fig:ideation_board}). Users can select, refine, or expand an option into new branches, with prior items preserved for comparison and traceability.
The Scripting board converges these ideas into structured outlines and scene lists (Fig. \ref{fig:scripting_board}), while the Storyboard integrates narrative and design into visual sequences; both adopt a revision-oriented model that foregrounds the latest draft while retaining earlier versions. 
The Design board manages visual assets such as characters and environments, organizing them in a tree structure for easy tracing and comparison.

When new blocks are generated, their stage boards auto-expand and center on the latest output, while older blocks can be collapsed to reduce clutter (Fig. \ref{fig:ideation_board}). Blocks consist of elements, enabling fine-grained interaction: users may refine a whole block (e.g., a scene list) or select specific elements (e.g., a single story beat or character design) for extension or refinement.

Together, these boards serve as the project canon, capturing divergence and convergence and ensuring continuity across the pipeline. This lineage-aware visualization lets creators trace how ideas evolve, capabilities absent in flat chat logs or generic file-based tools, while aligning with familiar creative practices.
The boards also serve as external stage memory for agents, providing a persistent record to reference when generating, refining, or revisiting outputs, reducing context load and supporting cross-stage consistency.

\subsubsection{Chat UI}
The chat panel manages user–agent interaction (Fig.~\ref{fig:ui}-A), with icons showing the active agent. Agents display status while processing (e.g., “Core Agent is thinking,” executing, or delegating) so users can track progress in real time. When a block or element is selected, it appears above the chat for context, and users can upload reference images (e.g., style inputs or sketches). While processing, new instructions are disabled, but users may interrupt with a Stop button, which restores the agent’s prior context and state.

\subsection{Technical Implementation}
\subsubsection{Runtime and Infrastructure}
AnimAgents is built on AutoGen Core 0.6.1~\cite{wu2024autogen} with routed agents. GPT-4.1-mini\footnote{GPT-4.1 mini, https://platform.openai.com/docs/models/gpt-4.1-mini}
powers most agent behaviors, while GPT-4.1-nano handles lightweight tasks like memory retrieval. The backend is implemented in Python with FastAPI and asynchronous orchestration, and the frontend in Next.js. A dedicated ReplyAgent bridges agent messages to the UI by pushing typed SSE events into per-session queues, enabling real-time streaming of progress and outputs.

\subsubsection{Memory and Context}


Each agent maintains short-term memory in a context window storing recent messages, tool calls, inter-agent communications, and outputs. The Core Agent also uses a long-term memory subsystem, following the indexing–retrieval–reading approach in~\cite{wu2025longmemeval}. We simplified variants of the indexing and retrieval strategies: recent interactions are periodically grouped into time-ordered chunks with concise summaries and indexed in a ChromaDB\footnote{ChromaDB, https://www.trychroma.com/}
vector store for later lookup. Before each session, the latest user message is expanded with lightweight query expansion to retrieve relevant past interactions via similarity search, which are then supplied to the Core Agent.

%

\subsubsection{Image Generation Pipeline}
For image generation, we integrate Flux.1-Kontext-pro\footnote{Flux.1-Kontext-pro, https://bfl.ai/models/flux-kontext}
, which supports unconditional and conditional generation with reference images via tool chaining. Prompts are built dynamically with stage-aware structure (e.g., hero images, character sheets, storyboard panels) and may include user references to preserve style and character consistency. Style frames and storyboard panels are generated in parallel for visual coherence. Images are stored locally, embedded into blocks, and displayed in the UI with other results.

\subsubsection{Error Handling and Robustness}
For error handling, each request carries a CancellationToken. Core and Specialized Agents trap cancellations or exceptions at delegation and response points, emit user-visible events with error details, and safely terminate execution. The agent context and system state then roll back to the prior session to preserve consistency.

In summary, AnimAgents realizes our three design goals. First, the Core and Specialized Agents coordinate multi-stage workflows and preserve consistency across narrative and visual outputs through stage-aware orchestration (DG1). Second, the multi-board canvas with lineage-aware blocks offers structured visualization aligned with artistic practice, letting creators organize, compare, and trace outputs more effectively than flat chat or file-based systems (DG2). Third, the Core Agent manages stage progression, interprets intent, and orchestrates agents that generate variations while supporting element-level refinement for targeted revisions, expanding creative possibilities while preserving user control (DG3). Together, these mechanisms demonstrate how multi-agent orchestration can be transformed into a human-facing interaction paradigm supporting end-to-end pre-production with clarity, continuity, and creative ownership.

\section{SUMMATIVE STUDY}

Our summative study evaluates how AnimAgents supports end-to-end animation pre-production. Rather than focusing only on efficiency or final output quality, an increasingly limited lens in the AI era, we examined how users interact with agents, how the system coordinates tasks, manages information, fosters creative agency and maintains continuity across stages.

\begin{figure}[htbp]
    \centering
    \includegraphics[width=1\linewidth]
    {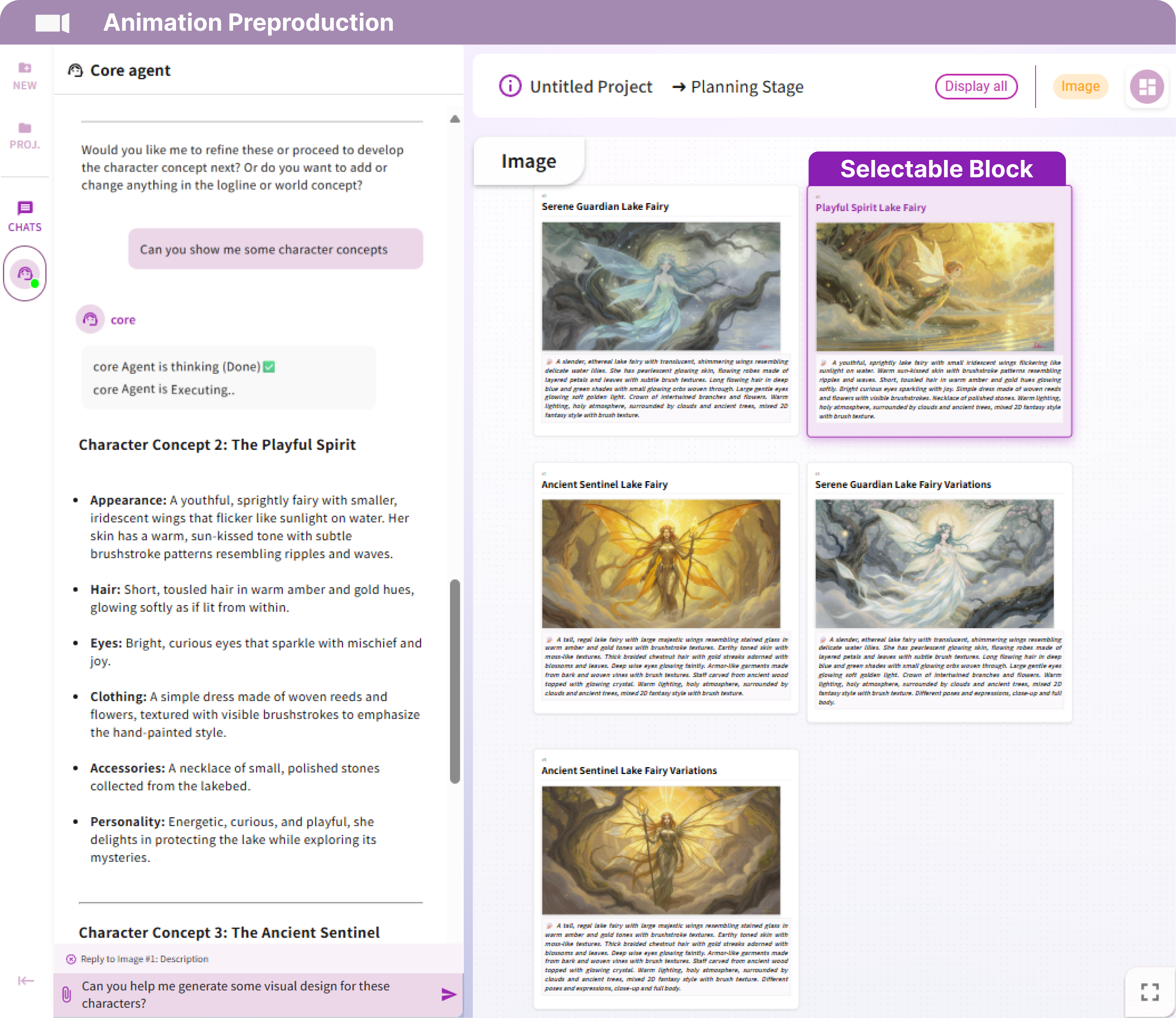}
    \caption{Baseline interface. Users interact with the agent via the chat panel (left). Visuals and descriptions generated during the conversation appear on the image board (right). Users can select an image block for further refinement.}
    \label{fig:baseline}
    \Description[This figure shows the baseline interface, with a chat panel for user–agent interaction and an image board for generated visuals that can be refined.]{This figure shows the baseline interface. On the left is a chat panel, where users interact with the agent by sending prompts and receiving responses. On the right is an image board, which displays visuals and image descriptions generated during the conversation. Each generated visual is presented as an image block. Users can select an image block to request further refinement or modification, supporting an iterative creative workflow. The interface emphasizes a simple two-panel layout: conversational input on the left and visual outputs on the right.}
\end{figure}

We conducted two complementary experiments:
\begin{itemize}
    \item \textbf{Controlled within-subjects task.} 
    We compared AnimAgents to a single-agent baseline (Fig. \ref{fig:baseline}). The baseline used the same AutoGen framework, LLM, and T2I model, guided by a tailored system prompt covering the full pre-production process. It supported conversational interaction, parallel image generation (e.g., all styleframes at once), and image selection and revision. Outputs appeared on a single board only—no stage separation, lineage tracking, or cross-stage organization—and non-visual content remained in chat. This baseline was intentionally stronger than typical industry workflows from our formative study, where creators separately used LLMs for ideation and T2I models for image generation, to control for automation and isolate AnimAgents' design contributions. This allowed us to test whether AnimAgents' multi-stage coordination and stage-specific visualization improved communication, consistency, and creative exploration.
    
    We did not compare against another MAS, as none support tailored human–AI collaboration for animation pre-production. Prior systems focus on automation and diverge too far from real-world creative practice to serve as a valid baseline.
    \item \textbf{An open-ended task.} Participants explored their own project with AnimAgents, and compared the experience to their previous workflow.
\end{itemize}

Our study aimed to answer the following research questions: 
\begin{itemize}
    \item RQ1(Coordination \& Continuity): Does AnimAgents improve stage coordination and narrative/style consistency, while reducing the time and steps required to complete pre-production tasks?
    \item RQ2(Organization \& Traceability): Does AnimAgents' stage-specific UI enhance output management, clarity, comparison, and version traceability to support creative decision-making?
    \item RQ3(Agency \& Creative Exploration): Does AnimAgents reduce unproductive overhead while better preserving user agency in creative exploration?
\end{itemize}

\subsection{Study Design}
\subsubsection{Participants}
We recruited 16 professional animation creators with 1-13 YoE, including 4 creative directors (P13-16) from three animation studios and 12 independent animators (P8-12, P17-23). 
5 participants (P8-12) had participated in an earlier formative study. All participants received ~\$75 USD compensation.

\subsubsection{Procedure}
Each study lasted about three hours and began with a short briefing. In the first task, participants completed a 15-minute tutorial and practice, a 40-minute pre-production task, and a 5-minute post-task questionnaire; system order and topics were counterbalanced. The second task was a 40-minute open-ended session where participants explored AnimAgents with their own project, followed by a comparative questionnaire. The study concluded with a 20–30 minute interview on both systems, their strengths and weaknesses, and AnimAgents' potential role in creative workflows.

\subsubsection{Task overview}
The first task mirrored the exploratory nature of pre-production. Participants worked on one of two exploratory topics: (1) Dream Architect and (2) The Living City, both designed for creative exploration. With each system, they moved from concept through visual design, story outline, and scene list, producing styleframes as the required deliverable, with optional storyboards if time allowed. To foster active collaboration over passive automation, participants were asked to create a short film concept they would personally want to watch, knowing final results would be rated. Tasks ended once styleframes were complete and participants were satisfied. Topics were reviewed and validated by a professional creative director (P3) from our formative study. Figure~\ref{fig:system_workflow} shows an example workflow for the first topic.

In the second open-ended task, participants recreated part of an ongoing or past project (e.g., short film concept, client commission) with AnimAgents, allowing direct comparison to their prior workflows (e.g., ChatGPT + MidJourney).

\subsubsection{Measurements}
\begin{figure*}[h]
    \centering
    \includegraphics[width=1\linewidth]{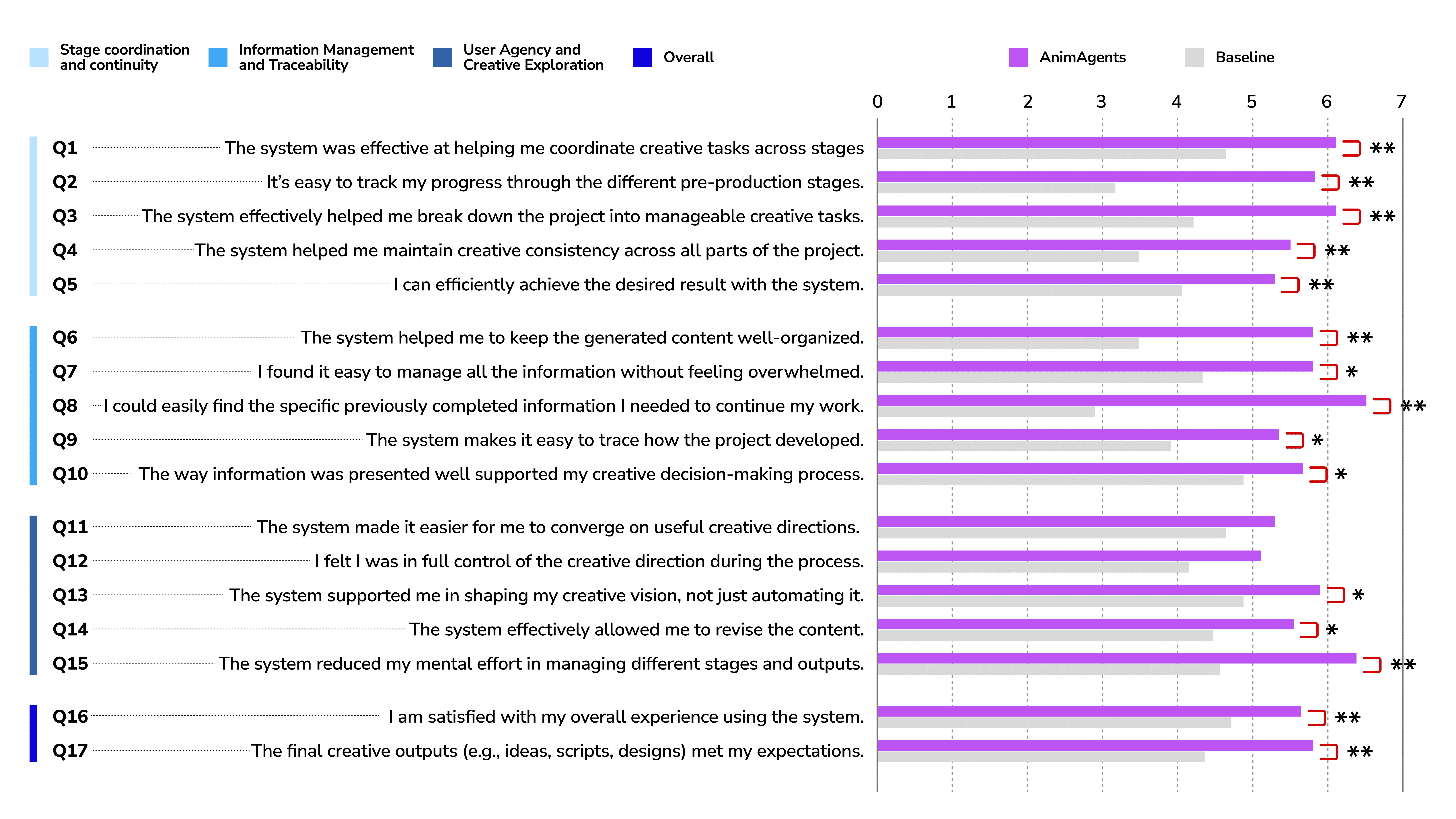}
    \caption{Survey results from the within-subject task. Participants rated Stage coordination and continuity, Information Management and Traceability, User Agency and Creative Exploration, and Overall satisfaction for both the Baseline and AnimAgents system using a 7-point Likert scale. *: \textit{p} < .05 and **: \textit{p} < .01.}
    \label{fig:result_within}
    \Description{This figure presents a comparative bar chart of participants’ responses to 17 survey questions about AnimAgents (purple bars) versus the Baseline (gray bars). Each question belongs to one of four categories: Stage coordination and continuity (Q1-Q5), Information Management and Traceability (Q6-Q10), User Agency and Creative Exploration (Q11-Q15), and Overall satisfaction (Q16-17), and is rated on a 7-point Likert scale (0 to 7). Along the vertical axis are the question labels, while the horizontal axis indicates average scores. The figure shows that AnimAgents scores higher than the Baseline on most questions, with 10 significant at p < .01 and 5 at p < .05; the remaining 2 show non-significant differences. Statistical significance denoted by a single asterisk (*) for p < .05 or a double asterisk (**) for p < .01.}
\end{figure*}

The post-condition questionnaire for the first task addressed our three research questions:  Stage Coordination \& Continuity (RQ1); Organization \& Traceability (RQ2); Agency \& Creative Exploration(RQ3). All questions and results appear in Figure~\ref{fig:result_within}. Responses were collected on a 7-point Likert scale (1 = strongly disagree, 7 = strongly agree). Following prior HCI and creativity research~\cite{lubos2024llm, wang2025gentune, palani2022don, son2024genquery}, we used a self-report approach and analyzed data with the Wilcoxon signed-rank test~\cite{woolson2005wilcoxon}. We also recorded task completion time and logged all user interactions, analyzing usage patterns by counting the frequency of each interaction type under both conditions.

\begin{figure*}[h]
    \centering
    \includegraphics[width=1\linewidth]{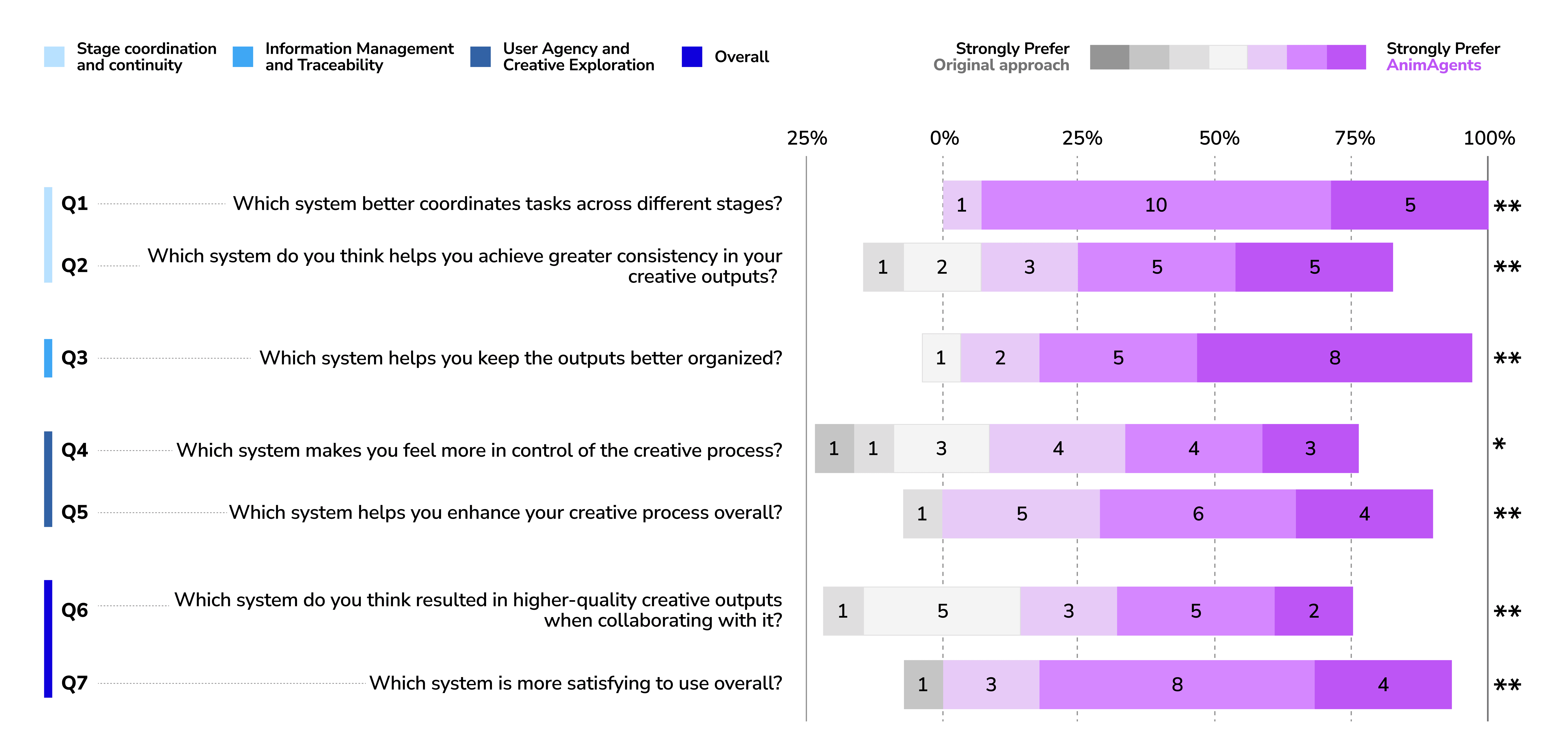}
    \caption{Survey results from the open-ended task. User rated their original approach and AnimAgents on a 7-point Likert scale across Stage coordination and continuity, Information Management and Traceability, User Agency and Creative Exploration, and Overall satisfaction. *: \textit{p} < .05 and **: \textit{p} < .01.}
    \label{fig:comparative}
    \Description{This figure shows participants’ preference distributions for their original approach (left) versus AnimAgents (right), as rated on a 7-point Likert scale. Each of the 7 questions (Q1–Q7) falls under one of four focus areas: Stage coordination and continuity (Q1–Q2), Information Management and Traceability (Q3), User Agency and Creative Exploration (Q4–Q5), and Overall satisfaction (Q6-Q7). The horizontal axis ranges from 25\% (slightly favoring the original approach) on the left, through 0\% (neutral preference), to 100\% (strongly favoring AnimAgents) on the right. Each bar’s segment length and the numbers displayed indicate how many participants’ ratings fall into that range. Overall, the data shows a clear shift toward AnimAgents across all questions, with 6 significant at p < .01 and 1 at p < .05.}
\end{figure*}

For the second open-ended task, the questionnaire asked participants to compare AnimAgents with their prior GenAI workflow across all research dimensions. Responses were collected on a 7-point Likert scale (1 = strong preference for the prior approach, 7 = strong preference for AnimAgents). Questions and results appear in Figure~\ref{fig:comparative}. We analyzed responses with a one-sample Wilcoxon signed-rank test against the neutral midpoint (4), appropriate for ordinal data from Likert-scale preference questionnaires~\cite{roberson1995analysis, capanu2006testing, taheri2013generalization}.

In-depth interviews complemented the controlled study by providing qualitative insights into baseline comparisons and how professionals applied AnimAgents to real projects. We examined its impact on our three research questions. Participants elaborated on questionnaire responses and reflected on system strengths, challenges, and potential improvements (see Appendix B). All interviews were transcribed and summarized. The initial coding was led by the first author with animation studio experience and refined with two additional researchers until consensus.

\section{RESULTS \& FINDINGS}

In this section, we report the findings organized by our three research questions (RQ1-RQ3), combining the results of both the controlled experiment and open-ended task.

\subsection{RQ1: Coordination \& Continuity}
\subsubsection{Improved coordination and task management}
AnimAgents significantly improved participants’ ability to coordinate pre-production creative tasks, track progress, and break down projects into manageable subtasks across stages compared to the single-agent baseline (Fig. \ref{fig:result_within}, Q1–Q3, \textit{p} < .01). In the open-ended task, 100\% of the participants preferred AnimAgents for cross-stage coordination (Fig. \ref{fig:comparative}, Q1). Many described AnimAgents more as a project manager than just a consultant, like Baseline. (P10–11, P13, P15–17, P19–22).

Participants highlighted AnimAgents' ability to coordinate each task: “\textit{It consolidates each stage’s task and systematically places everything from different departments into one place}” (P8). Others praised its flow management: “\textit{It clearly plans things step by step}” (P10). One participant reflected: “\textit{I often lose track of stages, but this system keeps the workflow clear}” (P22).
Task management also emerged as a strength: “\textit{It divides big tasks into smaller ones, which lets me focus on the task without missing something}” (P14). 

However, a few participants found the stage-based workflow too rigid: “\textit{The system felt too standardized and didn’t follow the workflow I wanted}” (P20). For example, one participant noted: “\textit{I usually start directly with storyboard sketches, but the system required me to finish the story first}” (P23).

\subsubsection{Maintaining creative consistency}
AnimAgents significantly improved creative consistency across stages and better met participants’ expectations (Fig. \ref{fig:result_within}, Q4, Q17, \textit{p} < .01). 81\% preferred AnimAgents for consistency (Fig. \ref{fig:comparative}, Q2). Participants were particularly impressed by the visual style consistency (P8, P12, P14–17, P21–23): “\textit{From character design to styleframes and storyboards, the style was maintained precisely}” (P22). Another explained: “\textit{When using other tools, the style would shift even with the same description in prompts. But with AnimAgents, it remained consistent.}” (P21).

Narrative consistency was also improved compared to the baseline: “\textit{With the first system (baseline), the story often drifted from the beginning. With AnimAgents, I could choose what to extend, and the storyline stayed consistent}” (P19). However, issues arose with longer text inputs. For example, one participant gave the Core Agent a full script and asked for a shot list, but the result diverged from the script: “\textit{I had to directly talk to the Scripting Agent instead}” (P15).

\subsubsection{Greater efficiency}
Although AnimAgents' multi-agent interactions sometimes took longer than the baseline, participants nonetheless rated it as significantly more efficient for achieving desired results (Fig. \ref{fig:result_within}, Q5, \textit{p} < .01). On average, tasks were completed faster with AnimAgents (Mean = 34.4) than with the baseline (Mean = 39.6), a reduction of 5.2 units ($\approx$13\%). A paired t-test confirmed this improvement (\textit{p} < .01). Participants primarily attributed these gains to improved traceability and coherence.

Three main factors were identified as sources of efficiency: (1) the ability to easily retrieve prior outputs (P8, P10, P14–15, P18, P20–21), (2) reduced inconsistency across results (P8, P10–11, P13, P15, P18, P22), and (3) clear step-by-step guidance (P11, P19–20). As one participant explained, “\textit{It’s easy to find previous information}” (P8). Others emphasized reduced frustration, for example: “\textit{I no longer need to keep fixing inconsistencies}” (P10). Clearer system guidance also improved pace: “\textit{The instructions were very clear}” (P20).


\subsection{RQ2: Organization \& Traceability}
\subsubsection{Output Management}
AnimAgents was rated significantly better than the baseline for keeping content organized and managing information without overload (Fig. \ref{fig:result_within}, Q6 \textit{p} < .01; Q7 \textit{p} < .05), with 94\% of participants preferring it (Fig. \ref{fig:comparative}, Q3). One noted, “\textit{I usually dislike organizing, but this system felt like it organized the ideas in my head for me}” (P16).

Participants highlighted the value of stage-specific multi-boards (P8–11, P14–15, P17–21). “\textit{It's clear what each block is for, so it’s easy to focus on that part of the process}” (P17). Another added, “\textit{The board layout and how each section is presented closely matches the logic of my past work, very intuitive}” (P13). The block structure also reduced overload (P8, P10, P14–15, P18, P20–21): “\textit{The UI makes it clear what was generated and lets me compare outputs. Unlike the baseline, where I lost track of ideas as the session grew}” (P11). Participants further appreciated that “\textit{The system only produces requested content, so the information feels more concentrated}” (P20).

However, participants noted that too many blocks could be confusing across stages: “\textit{It’s easy to mix up blocks from different stages}” (P21). Others suggested adding block deletion (P9, P11, P14, P22–23) or preferred a unified view: “\textit{The whole canvas could be a single board, with each node representing a different board type—this would make the relationships clearer}” (P20).


\subsubsection{Traceability}
AnimAgents outperformed the baseline in helping participants locate prior outputs and trace project development (Fig. \ref{fig:result_within}, Q8 \textit{p} < .01; Q9 \textit{p} < .05). “\textit{It’s much easier to find earlier information, and everything is already categorized}” (P9). The board-based interface matched participants’ practices: “\textit{My brain works visually, so I remember where each block is}” (P20); “\textit{It feels like how I usually organize reference images}” (P12). Many contrasted this with chat-only tools (P8–12, P19, P22): “\textit{With ChatGPT, I just kept scrolling up and slowly searching for what I needed}” (P14).

\subsubsection{Creative decision}
Participants further reported that AnimAgents' organization and traceability supported creative decisions (Fig. \ref{fig:result_within}, Q10 \textit{p} < .05). One explained, “\textit{This organized system helps me better structure ideas and make, know where mistakes happen}” (P21). Stage-specific boards were especially helpful (P9–11, P13, P15, P17, P19–20, P22–23). “\textit{The tree structure and options in the Ideation stage made it easier to see how ideas evolved compared to GPT}” (P15); “\textit{The styleframe and storyboard boards let me quickly compare text and visuals}” (P17).

\subsubsection{Communication Facilitation}
Finally, participants suggested that AnimAgents' organized outputs could facilitate communication with both collaborators and clients (P8–11, P14–15, P18–19, P21–23). For internal use, “\textit{Collaborating directors can see the entire pre-production thought process}” (P8).
Another added, “\textit{The information flow is complete. Everyone can understand the reasoning behind the design.}” (P13).
For external communication, participants compared the system’s boards to professional presentation materials: “\textit{You can directly present it to clients and demonstrate different versions on the spot}” (P23).


\subsection{RQ3: Agency \& Creative Exploration}
\subsubsection{Controllability}
AnimAgents was rated significantly better than the baseline in effectively revising outputs (Fig. \ref{fig:result_within}, Q14, \textit{p} < .05). Many praised element-level selection (P9–12, P17, P19, P21–23). As one explained, “\textit{With other AI tools, I had to regenerate everything, which was chaotic. AnimAgents lets me adjust only the selected part, much clearer}” (P18). Another noted, “\textit{I can isolate and edit just the part I want, while keeping the rest unchanged, it feels like working with layers}” (P9).

It also better supported shaping creative vision rather than automating tasks (Fig. \ref{fig:result_within}, Q13, \textit{p} < .05). Users appreciated that AnimAgents followed their instructions faithfully without adding unnecessary content (P8, P11–12, P14, P19, P21–23): “\textit{AnimAgents does exactly what I ask, unlike ChatGPT}” (P17). Another noted, \textit{“Extra outputs distract me. With AnimAgents, my creative goals stay on track}” (P14).

Ratings for full creative control were not significant (Fig. \ref{fig:result_within}, Q12, \textit{p} = .06), often due to Core Agent occasional misjudgments in delegation, likely from LLM variability. For example, “\textit{I wanted a hero image, but it gave me a character design}” (P16). Others noted vague prompts could be misinterpreted: “\textit{If I just say ‘I want the story,’ it might treat it as a story outline in Scripting or a concept in Ideation}” (P19).

Despite these issues, 69\% still preferred AnimAgents for providing a stronger sense of control (Fig. \ref{fig:comparative}, Q4), due to the way the Core Agent communicated with them: “\textit{It never makes decisions on its own, it always asks for my approval}” (P11); “\textit{It guided me without interfering too much with my creativity. I felt in control}” (P21).

\subsubsection{Creative exploration}
Compared to prior GenAI workflows, 94\% preferred AnimAgents for enhancing the overall creative process (Fig. \ref{fig:comparative}, Q5). However, ratings for supporting convergence on useful directions were not significant (Fig. \ref{fig:result_within}, Q11, \textit{p} = .10). The participants noted that both systems improved creativity but differently (P8, P11-12, P14-17, P19, P21-24). The baseline offered more variety but less focus: “\textit{It gave me more ideas, but it was easy to lose track}” (P19). AnimAgents improved convergence yet sometimes constrained exploration: “\textit{AnimAgents organizes my thoughts when scattered, but if I want more ideas, I’d use the other system}” (P19).

Finally, AnimAgents was rated significantly higher for reducing mental effort and overall satisfaction (Fig. \ref{fig:result_within}, Q15–16, \textit{p} < .01), with 94\% preferring it on satisfaction (Fig. \ref{fig:comparative}, Q7). Participants highlighted its ability to filter and streamline information (P9–11, P13, P16–18, P22–23): “\textit{ChatGPT would dump everything, including lots of useless information. With AnimAgents, I don’t waste time reading it}” (P10); “\textit{It filters out unnecessary information for me}” (P11). 



\begin{table*}[t]
\centering
\small
\begin{threeparttable}
\begin{tabular}{lcccc}
\toprule
\textbf{Metric} & \textbf{Baseline} M$\pm$SD & \textbf{AnimAgents} M$\pm$SD & \textbf{$\Delta$} M [Md] & \textbf{\% $\uparrow$ (n/N)} \\
\midrule
Total messages     & 22.00$\pm$7.47 & 27.12$\pm$7.96 & +5.13 [5]  & 88\% (14/16) \\
Exploratory        & 10.69$\pm$3.65 & 12.25$\pm$5.34 & +1.56 [2]  & 81\% (13/16) \\
Revise             & \phantom{0}8.88$\pm$4.05 & \phantom{0}8.19$\pm$3.94 & $-$0.69 [$-$1] & 44\% (7/16) \\
Confirmation       & \phantom{0}1.75$\pm$1.81 & \phantom{0}2.56$\pm$1.67 & +0.81 [1]  & 56\% (9/16) \\
Directive          & \phantom{0}2.31$\pm$1.78 & \phantom{0}6.00$\pm$3.41 & +3.69 [4]  & 81\% (13/16) \\
\addlinespace
\multicolumn{5}{l}{\textit{Pooled shares of messages (\%) across all runs}} \\
\midrule
Exploratory (share)  & \multicolumn{1}{c}{48.6} & \multicolumn{1}{c}{45.2} & \multicolumn{2}{c}{} \\
Revise (share)       & \multicolumn{1}{c}{40.3} & \multicolumn{1}{c}{30.2} & \multicolumn{2}{c}{} \\
Confirmation (share) & \multicolumn{1}{c}{\phantom{0}8.0} & \multicolumn{1}{c}{\phantom{0}9.4} & \multicolumn{2}{c}{} \\
Directive (share)    & \multicolumn{1}{c}{10.5} & \multicolumn{1}{c}{22.1} & \multicolumn{2}{c}{} \\
\bottomrule
\end{tabular}
\caption{Combined log analysis: per-run message behavior and pooled shares (N=16 paired runs). $\Delta$ is AnimAgents minus Baseline.}
\label{tab:msg_combined}
\Description[This table compares message behaviors between Baseline and AnimAgents across 16 paired runs, showing totals, categories, and pooled shares.]
{This table presents a combined log analysis of messaging behavior in 16 paired runs, comparing Baseline and AnimAgents.  
For total messages, participants sent on average 22.00 (SD 7.47) with the Baseline and 27.12 (SD 7.96) with AnimAgents, an increase of about 5 messages, with 14 of 16 runs (88\%) showing higher counts under AnimAgents.  
Exploratory messages rose from 10.69 (SD 3.65) to 12.25 (SD 5.34), an increase of about 1.6 messages, with 81\% of runs showing gains. Revise messages decreased slightly from 8.88 (SD 4.05) to 8.19 (SD 3.94), with fewer than half the runs (44\%) showing increases. Confirmation messages grew from 1.75 (SD 1.81) to 2.56 (SD 1.67), while directive messages showed the largest gain, rising from 2.31 (SD 1.78) to 6.00 (SD 3.41), with 81\% of runs showing increases.  
The pooled message shares indicate a redistribution across categories. Exploratory messages made up 48.6\% of all messages in the Baseline and 45.2\% with AnimAgents. Revise messages dropped from 40.3\% to 30.2\%. Confirmation increased slightly from 8.0\% to 9.4\%. Directive messages more than doubled, from 10.5\% in the Baseline to 22.1\% with AnimAgents. 
These results highlight that while overall messaging increased, the most notable shift was a strong rise in directive messages, alongside a relative decline in revising behavior. All underlying mean and standard deviation values are provided in the table for reference.}
\end{threeparttable}
\end{table*}

\subsection{Analysis of User Interaction}
To analyze participant interactions, we classified each message into four categories: Revise (modifying or refining existing content, e.g., “\textit{Remove the accessories on the character}”); Confirmation (accepting results without adding constraints, e.g., “\textit{Yes, let’s keep this version}”); Exploratory (requesting new content or ideas, e.g., “\textit{Give me five story concept options}”); and Directive (providing workflow-level or meta-instructions, such as “\textit{Next, let's focus on the character's visual}” or “\textit{The art style is 2D picture book, for kids}”).

Log analysis revealed distinct patterns (Table~\ref{tab:msg_combined}). Directive interactions showed the largest increase with AnimAgents (22\% vs. 10\%), rising for 13 of 16 participants and suggesting greater engagement in workflow steering and delegation. Confirmations also rose slightly, as users signed off on outputs more often. In contrast, Revisions decreased (30\% vs. 40\%), indicating fewer corrective edits. Exploratory interactions remained proportionally stable but higher in absolute count. Overall, the logs suggest that AnimAgents shifted the interaction balance toward higher-level control: less correction, more direction, and approval of the process, reflecting its role as a project manager–like collaborator.

Overall, AnimAgents offered clear benefits over the single-agent baseline in stage coordination, output management, and user agency. Participants valued its project manager–like support for organizing tasks and maintaining consistency, stage-specific boards that improved organization and traceability, and element-level revision tools that aligned outputs with intent. While some noted rigidity, Core Agent misinterpretation, or reduced divergence, the system consistently lowered cognitive effort, increased satisfaction, and helped participants stay in control of the multi-stage creative process.

\section{FIELD STUDY}

We conducted a week-long field study with 4 professional animation creators to examine how AnimAgents integrates into real-world pre-production. The study explored how practitioners adopt the system across different contexts and how these contexts shape interaction patterns and perceived value. Specifically, we asked:
RQ4: How do professionals integrate AnimAgents into their ongoing animation pre-production workflows, and what contexts shape its perceived value?

\subsection{Participants, Study Procedure and Evaluation}
We recruited 2 creative directors and 2 independent animators for the field study (Table~\ref{table:field}). P5 (age 37, 12 YoE) is from a major commercial studio with 145 employees, and P24 (age 36, YoE) from a local original animation studio with 23 employees. P8 (age 26, 2 YoE) and P19 (age 32, 8 YoE) are freelance animators working on commissioned character projects. P8 and P19 had joined our summative study and volunteered for field deployment, while P5 and P24 were recruited through direct outreach, including a system demo and interview. All four integrated AnimAgents into their ongoing workflows during the study.

We deployed the same AnimAgents system from the summative study on a web server. Participants integrated it into ongoing pre-production tasks and kept diaries documenting stage coordination, interactions, transitions, and use of outputs. Afterward, we conducted 60-minute interviews on AnimAgents' workflow integration, interaction experience, and perceived benefits and limitations. The protocol followed the same dimensions as the research questions in our summative study. Due to NDAs, we could not publicly share the studios’ final results.

\begin{table}[h]
\centering
\small
\begin{tabular}{|c|c|c|}
\hline
\textbf{ID} & \textbf{Project Type} & \textbf{Current Progress} \\ \hline
P5  & Product Commercial Animation        & Scripting        \\ \hline
P8  & Individual Character Animation      & Exploratory   \\ \hline
P19 & Individual Character Animation      & Early Ideation      \\ \hline
P24 & Original Animated Series             & Mid Ideation     \\ \hline
\end{tabular}
\caption{Project types and current pre-production progress for field-study participants.}
\label{table:field}
\Description[This table lists project types and progress for four field-study participants.]{This table is organized into three columns: ID, Project Type, and Current Progress, with one row for each participant. It presents the project types and current pre-production progress of four participants in the field study. Participant P5 worked on a product commercial animation that had reached the scripting stage. Participant P8 focused on an individual character animation project and was in the exploratory stage. Participant P19 also worked on an individual character animation but was still in early ideation. Participant P24 developed an original animated series, which was in the mid-ideation stage.}

\end{table}

\subsection{Result and Finding}

\subsubsection{Commercial vs. Original Animation Studio}
Both directors emphasized AnimAgents' ability to produce coherent storyboards and styleframes that preserved visual consistency, accelerating communication with collaborators. P24 noted, “\textit{Consistency across styleframes and storyboards was impressive: once I uploaded reference sheets, the results matched surprisingly well}.” P5 added, “\textit{AnimAgents quickly gave me several complete storyboard versions. Even if not every frame was usable, it was more than enough to communicate with clients}.”

Yet, the two directors’ experiences diverged sharply by production context. In client-driven commercial projects with tight deadlines, AnimAgents was valued for speed and efficiency. P5, working on a one-minute laptop promotion, called it “\textit{much faster than manually piecing together shots in 3D software},” noting that parallel, coherent generation let him deliver multiple storyboard versions in a day instead of a week. He summarized its benefits as “\textit{consistency, efficiency, and ease of revision},” emphasizing AnimAgents' fit for fast-turnaround product animation.

In original animation, where stories evolve across episodes and directors demand high narrative and visual standards, AnimAgents played a more ambivalent role. P24 valued its project management support: “\textit{I felt like a project manager guiding me, so I could focus on story direction.}” Yet she stressed its limits. Narratively, the outlines lacked causal depth: “\textit{We quickly ran into logical problems, for example, what event could realistically cause a character’s dark turn?}” and “\textit{AI doesn’t really break the frame, it rarely gives you something unexpected.}” Visually, she was even more critical: “\textit{Original projects have extremely high detail requirements. Directors insist on full consistency. AI visuals rarely work.}” While styleframes sometimes sparked discussion, she concluded, “\textit{for unique artistic styles, AI is at best a bonus reference, not a production asset.}” 

These cases demonstrate that while AnimAgents provides consistency and structured outputs, its value diverges by context: commercial directors prioritize efficiency and client-ready deliverables, while original directors view it mainly as support for organizing ideas and rough references, falling short in terms of depth, originality, and visual rigor.

\subsubsection{Supporting Independent Animators}
For freelancers, AnimAgents' value was in scaffolding pre-production steps they often rushed or skipped. Both P8 and P19 are 3D animators who usually rush pre-production to begin 3D work. P8 had used GenAI mainly for 2D ideation, while P19 sketched only rough storyboards. Both found AnimAgents especially useful for scripting, styleframes, and storyboards, steps that they typically skipped or compressed.

\begin{figure*}[t]
    \centering
    \includegraphics[width=\linewidth]{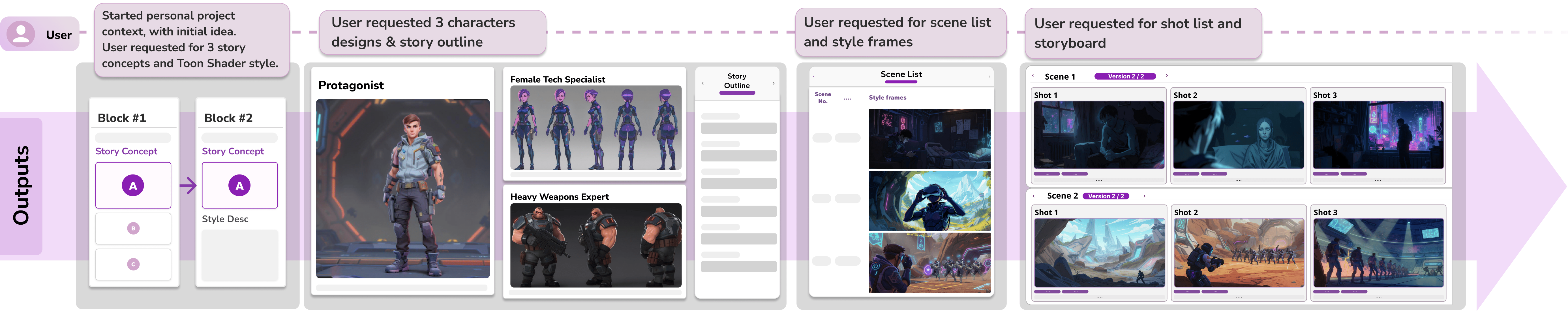}
    \caption{P8’s workflow and results with AnimAgents in the field study. Starting from an initial idea, he first asked for story concepts, then moved on to character designs and a story outline. He then progressed to a scene list, styleframes, and ultimately a shot list and storyboard.}
    \label{fig:fieldstudy}
    \Description[This figure illustrates P8’s workflow and outputs using AnimAgents.]{This figure shows P8’s step-by-step workflow with AnimAgents during the field study, represented in a left-to-right sequence of panels. 
    On the far left, a “User” icon indicates that he began a personal project by providing context and an initial idea, requesting three story concepts in a Toon Shader style. 
    The next column shows two blocks labeled “Block \#1” and “Block \#2,” each containing a selected story concept and a style description. 
    In the following section, three character design outputs are displayed: a male protagonist, a female tech specialist, and a heavy weapons expert, alongside a generated story outline. 
    Moving further right, a “Scene List” panel presents several visual style frames, including a dark room with the protagonist’s sick mother, the protagonist wearing a VR helmet to enter the arena, and a battle sequence featuring the protagonist. 
    Finally, on the far right, the storyboard panel depicts two scenes, each with three shot images that visually detail the narrative progression. 
    Arrows and labels across the top summarize the user’s requests at each stage, highlighting the iterative expansion from abstract ideas to concrete pre-production assets.}

\end{figure*}
Figure~\ref{fig:fieldstudy} (Appendix C) shows P8’s collaboration with AnimAgents on a personal project inspired by Ready Player One. As P8 noted, AnimAgents let him “\textit{move quickly from one step to the next,}” turning an initial idea into integrated narrative, design, and storyboard materials. Starting with story concepts, the system guided premise development, character design, and shot planning, producing coherent styleframes and a storyboard. While simplified here, his actual workflow involved multiple iterations and branches. Reflecting, he shared: “\textit{The system gave me an intuitive impression, organized everything quickly, and kept reminding me what to do next. When I asked it to expand a story in film or animation style, the results were often interesting. It helped me move quickly to the next step, and both the stories and visuals were very usable as references.}” He later used these directly in production.


P8 also explored the role-playing capability of AnimAgents, an unexpected and inspiring use case, by assigning creative personas to agents: “\textit{When I asked the Ideation Agent to think like Nolan, the director of Inception, it gave me non-linear narratives and montage-style transitions.}” This helped him explore directions beyond his usual thinking.
For P19, AnimAgents' value lay in project management and clarity. She described it as “\textit{like a little assistant: I could act as a creative director, assigning tasks while supervising results}.” Structured boards and flowcharts made revisions clear: “\textit{Each stage was neatly organized with diagrams to compare changes, so the process never felt messy. The adjustments worked smoothly.}”

In sum, independent animators valued AnimAgents for scaffolding pre-production steps they previously minimized, with P8 emphasizing its creative role-play potential and P19 its organizing and project-management support.


\section{DISCUSSION, LIMITATIONS, AND FUTURE WORK}
\subsection{Adapting Multi-Agent Systems to Nonlinear Creative Pipeline}
Our findings highlight three central design challenges in adapting multi-agent systems (MAS) to nonlinear creative workflows. Each reveals tensions between system design and creative practice, with implications for future creativity support tools.

\subsubsection{Nonlinear Workflows and Stage Flexibility}
Multi-stage, multi-agent systems(MAS) can coordinate cross-stage outputs without overloading a single agent~\cite{pan2025agentcoord}. Many current implementations remain linear, stage after stage, in science~\cite{schmidgall2025agent} and creative production~\cite{shi2025animaker, xu2024filmagent}. However, creative workflows are rarely linear; they are dynamic and iterative~\cite{sawyer2013zig, botella2011dynamic, botella2019dynamic}, and creators routinely revisit and refine earlier decisions~\cite{baldwin1962creative, fox2019mind, schuldberg2021creativity}. In animation pre-production, for example, story structure may be revised during storyboard or sketch boards before scripting.

This nonlinearity creates a key design tension: agents must handle out-of-order requests while maintaining global awareness, risking overload and misalignment ~\cite{beckenbauer2025orchestrator}. AnimAgents addresses this with stage-aware orchestration—interpreting intent, switching (or letting users switch) stages, and executing tasks in context. This reduces cognitive overhead and improves alignment, but cannot cover all workflow deviations. For example, When revising a character design late, AnimAgents switched to Design but failed to update earlier story decisions, requiring manual fixes. By contrast, single-agent systems (our baseline) are more flexible and adapt readily to unexpected changes~\cite{weiss1995adaptation, parkes1997learning}, but sacrifice cross-stage coherence and traceability. Balancing flexibility with structured consistency remains a central design challenge for AI in creative workflows. Future systems could let users toggle between project-manager guidance and free-form exploration, or sync agents when jumping across stages.

\subsubsection{Instability of LLM-based Multi-Agent Systems}
Another key challenge is the instability of LLM-based multi-agent systems: specification errors, inter-agent misalignment, and weak verification are common~\cite{cemri2025multi}. In creative settings, such errors can misinterpret intent or misroute tasks, and since processes are time-consuming, mistakes are costly and reduce perceived control. Stronger models lower error rates but increase latency, creating a speed–reliability trade-off. AnimAgents currently allows users to invoke specific agents as a corrective fallback, yet the Core Agent sometimes misrouted tasks, assumed correctness, and became trapped in repetitive loops. This shifted the burden of correction onto users and added cognitive load. 

These findings suggest reliability should be supported proactively, not repaired retroactively. Future versions could add lightweight validation agents or self-consistency checks on Core Agent decisions; though adding interaction time, these could improve reliability and reduce rework, making MAS more dependable in creative practice.

\subsubsection{Long-Horizon Memory and Traceability}
Creative projects often span weeks or months, requiring continuity across a large number of artifacts and decisions. Yet even with retrieval-augmented generation (RAG), LLMs frequently fail to maintain long-term context~\cite{maharana2024evaluating, zhang2024survey}, leading to inconsistencies across stages.

AnimAgents addressed this by externalizing intermediate artifacts as addressable blocks on stage-specific boards. Users could select, branch, and revisit outputs, creating stable project memory and traceable provenance that supported coherence and helped them reorient after interruptions. However, our memory implementation still sometimes destabilized the agent performance, as retrieval did not always surface the correct context~\cite{agrawal2024mindful, joren2024sufficient}. Future systems could strengthen retrieval methods or develop inter-agent documentation, where agents collaboratively maintain shared logs to ensure more reliable long-horizon memory.

In sum, adapting MAS to creative workflows requires balancing flexibility, reliability, and long-term traceability. AnimAgents offers initial solutions but also reveals design tensions that future CST research must address.

\subsection{From Animation to Broader Domains}
The core concept of AnimAgents is a stage-aware orchestration architecture that enables human–multi-agent collaboration in complex multi-stage workflows. Our work addresses a key challenge in current AI-assisted pipelines: supporting users in maintaining coherence and avoiding information overload, while also preserving creative agency when managing multiple GenAI tools~\cite{zhang2025research, yatani2024ai, patwardhan2024automated}.
This challenge extends to narrative media fields assisted by GenAI, such as film, TV programs~\cite{zhang2025generative, kalinski2025addressing}, and game development~\cite{begemann2024empirical, sun2023language, li2024unbounded}, which, like animation, require consistent coordination in the narrative, visual, and design processes. 

MAS naturally coordinate outputs from heterogeneous sources~\cite{sun2025multi}. They also mirror real-world collaboration: distributing tasks among specialized agents reflects professional role divisions and provides users with interpretability by indicating which “expert” to consult~\cite{sapkota2025ai}. This role-based delegation is not unique to animation but generalizes to other creative and knowledge-intensive domains~\cite{tran2025multi, schombs2025conversation}. AnimAgents' mechanisms, stage-specific boards, block-based external memory, and a project manager-like Core Agent, can be further adapted to these settings. Its modular design further supports other complex, non-linear workflows: new agents and tools can extend orchestration beyond pre-production into later stages, leveraging earlier assets for downstream tasks.



\subsection{Toward Real-World Creative Adoption}
Extensive research has explored real-world adoption of multi-agent systems in finance, healthcare, software engineering, and education~\cite{tran2025multi, xiao2024tradingagents, li2024exploring}, but far less in creative industries. Thus far, AnimAgents has only been tested in small, short-term deployments, while animation pre-production often spans weeks or months. While these studies demonstrate feasibility,  large-scale adoption remains untested. Long-term use may raise challenges such as sustaining consistency over months, adapting to evolving roles, and integrating with heterogeneous toolchains.

A key factor for adoption is integration into current practice. Our field study reveals that in commercial projects, where efficiency, iteration, and clear communication are crucial, GenAI systems are particularly valued, aligning with previous work~\cite{wang2025gentune, porquet2025copying}. AnimAgents accelerates iteration by coordinating outputs across stages and complements skills individual creators, particularly independent animators, may lack, echoing trends seen among content creators~\cite{anderson2025making, lyu2024preliminary, he2025recall}.
However, in original animation, directors often demand more detail and narrative depth than current AI can provide. What is needed is not more generated content but “\textit{information management, because projects span long periods and involve overwhelming amounts of information}” (P24). 

Unlike many GenAI tools that prioritize producing more or “better” outputs, AnimAgents was designed primarily as project management support. By strictly following user instructions and providing efficient orchestration, it preserved agency and adapted to both contexts, accelerating iteration in commercial projects and supporting coordination in original productions. This suggests a broader lesson for CSTs: enhancing coordination and control can matter more than improving generative quality. Future systems could adopt project-manager roles that facilitate teamwork across departments~\cite{han2024teams}, addressing real creative needs. For instance, contributors could upload outputs to shared boards for AnimAgents to organize, with all members retaining editing rights and visibility, while in co-directing scenarios it could bridge communication by making each director’s contributions visible to the other.

\subsection{Ethical Concern}
The increasing use of GenAI in animation~\cite{tang2025generative, gunanto2025future, clocchiatti2024character}, film~\cite{zhang2025generative, deng2024governance}, and games~\cite{begemann2024empirical, gallotta2024large} cuts time and costs but raises concerns about role displacement and pressure on artists, designers, and scriptwriters~\cite{gao2023aigc, bender2025generative, amankwah2024impending, caporusso2023generative}, and unresolved issues of intellectual property and unconsensual use of artists’ work~\cite{porquet2025copying, asperti2025critical}.
AnimAgents seeks to mitigate these risks by supporting exploratory processes for creative directors rather than producing final outputs. Yet in projects with lower aesthetic or narrative demands, participants noted its efficiency and control “\textit{could lower many entry barriers, and in the future many people could use such tools to create animation}” (P5). While this accessibility may aid independent creators, it also raises concerns about cost-conscious clients substituting roles~\cite{xiao2025exploratory, coetzer2025impact}.
Greater efficiency may also increase workflow pressure between departments and in client interactions~\cite{brynjolfsson2025generative}.
While AnimAgents positions itself as a workflow coordinator rather than a creativity substitute, broader risks persist. Overreliance on GenAI may reduce group creativity~\cite{kumar2025human, doshi2024generative} and homogenize artistic styles~\cite{asperti2025critical, porquet2025copying}, highlighting the need for continued research on the ethical and cultural impacts of human–AI co-creation.



\subsection{Limitation and Future Work}
Beyond the points mentioned above, this work has several limitations that suggest directions for future research.

\textbf{Study design.} Our evaluation relied mainly on self-report questionnaires. Although we recruited professionals and collected rich qualitative feedback, future studies could add external expert reviews of participant outputs to more rigorously assess AnimAgents effectiveness.

\textbf{Centralized orchestration.} AnimAgents lacks inter-agent communication. This reduces overhead but sometimes misdirects tasks. Future extensions could test “group meetings,” where agents exchange updates to improve coordination or engage in creative discussions with users.

\textbf{Cross-board relation.} While AnimAgents' block-based boards efficiently support traceability, cross-board lineage becomes challenging. Future designs could provide richer lineage visualization and direct trace-back functions, or more flexible lineage arrangement, enabling users to quickly identify how outputs originated across stages.

\textbf{Potential for role-play.} Beyond coordination, multi-agent systems can also role-play to stimulate creativity. By adopting complementary perspectives, or impersonating well-known figures, as in P8’s field study, agents may better align with creators’ intent and expand the design space for independent animators. Future work could explore role-based orchestration as a catalyst for human–multi-agent co-creation.

\section{CONCLUSION}
We introduced AnimAgents, a human–multi-agent system that coordinates animation pre-production by mirroring studio role structures. The Core Agent acts as a project manager, orchestrating specialized agents, tracking progress, and ensuring cross-stage consistency while reducing management overhead. Stage-specific boards provide a global view of the workflow, manage outputs, and preserve traceability, serving as external memory for both users and agents. Block- and element-based interactions further support divergent exploration and precise refinement. In a summative study with 16 professional animators, AnimAgents significantly improved coordination, information management, and creative agency compared to a strong single-agent baseline. A follow-up field study with creative directors and independent animators further demonstrated its potential to enhance real-world pre-production workflows.

\begin{acks}
This work was supported by the National Science and Technology Council
\end{acks}




\lstset{
  basicstyle=\ttfamily,        
  breaklines=true,             
  escapeinside={(*@}{@*)},     
  columns=fullflexible         
}

\appendix
\small\ttfamily
\label{Appendix_Formative}
\section{Appendix A: Interview Questions for Formative Study}
\subsection{Creative Directors}
\begin{lstlisting}
Workflow
1. Could you describe your overall pre-production workflow from start to finish?
2. Could you share some real examples from past projects during this stage, including the resources you used and intermediate artifacts (e.g., sketches, brainstorming documents)?
3. In your position, what were your main responsibilities?
4. How do you develop the overall creative concept for a project? What are the key steps?
5. What tools do you use during the process, and for what purposes?
6. How do you assign tasks to different teams?
7. Could you describe your collaboration process with different teams?
8. How do you review and approve the results delivered by your teams?

*Preproduction includes concept development, story, script, character and scene design, styleframes, and storyboarding.

Pain Points
1. Overall, what do you find the most difficult part of this workflow?
2. What kinds of resources or support would you most like to have?
3. When working with different team members, what are the biggest challenges?
   - For example, aligning everyone's opinions or repeatedly guiding them through the process.

AI Related
1. What is your general attitude toward using generative AI in the creative process?
2. Have you used AI tools in your pre-production workflow?
3. If yes, where and how did you use them?
4. What do you think are the main challenges of using AI tools in pre-production today?
5. If you could design an AI tool for pre-production, what should it be able to do, and what should it avoid?

AI Assistant
1. If you had an AI assistant to help you, what kind of interaction would you want with it?
2. Imagine a whole generative AI team with members responsible for ideation, scriptwriting, character/scene design, and visual development. You could direct them to complete creative tasks together (e.g., ideation expands the story, design produces multiple character versions, script/visual development generate styleframes and scenes). I will show some examples.
   - What are your thoughts on this scenario?
   - How do you think such a system could help you?
   - What strengths and weaknesses do you see?
3. How would you want to explore creativity with such a system?
4. How would you want to interact with the different AI members? For example, would you prefer to have a single assistant that coordinates everything, or communicate directly with each member?
5. How much control would you want over this team? Would you prefer to issue commands through natural language, or directly edit their outputs?
6. How would you use the results produced by this AI team, such as those in the example?
7. (CD) Do you think such a system could also help you communicate with your real-world team?
\end{lstlisting}

\subsection{Independent Animators}
\begin{lstlisting}
Workflow
1. Could you briefly describe your overall creative workflow from start to finish?
2. Could you describe your pre-production workflow from start to finish?
3. Could you share some real examples from past projects during this stage, including the kinds of resources you used and intermediate artifacts (e.g., sketches, brainstorming documents)?
4. How do you develop the overall creative concept for a project? What are the key steps?
5. What tools do you use during the process, and for what purposes?
6. How do you organize yourself to accomplish different tasks simultaneously?
7. What do you consider the most important part of pre-production?

*Preproduction includes concept development, story, script, character and scene design, styleframes, and storyboarding.

Pain Points
1. Overall, what do you find the most difficult part of this workflow?
2. What kinds of resources or support would you most like to have?
3. Do you encounter technical challenges in the workflow, such as wanting to accomplish something but lacking the skills? How do you overcome these?
4. Do you experience creative blocks, and if so, how do you resolve them?

AI Related
1. What is your general attitude toward using generative AI in the creative process?
2. Have you used AI tools in your pre-production workflow?
3. If yes, where and how did you use them?
4. What do you think are the main challenges of using AI tools in pre-production today?
5. If you could design an AI tool for pre-production, what should it be able to do, and what should it avoid?

AI Assistant
1. If you had an AI assistant to help you, what kind of interaction would you want with it?
2. Imagine a whole generative AI team with members responsible for ideation, scriptwriting, character/scene design, and visual development. You could direct them to complete creative tasks together (e.g., ideation expands the story, design produces multiple versions of characters, script/visual development produce styleframes and scenes). I will show some examples.
   - What are your thoughts on this scenario?
   - How do you think such a system could help you?
   - What strengths and weaknesses do you see?
3. How would you want to explore creativity with such a system?
4. How would you want to interact with the different AI members? For example, would you prefer to have a single assistant that coordinates everything, or communicate directly with each member?
5. How much control would you want over this team? Would you prefer to issue commands through natural language, or directly edit their outputs?
6. How would you use the results produced by this AI team, such as those in the example?
7. Do you think such a system could act as a virtual creative team under your leadership? What kinds of support would you most need from it?
\end{lstlisting}

\section{Appendix B: Interview Questions for Summative Study}
\label{Appendix_Interview}
\small\ttfamily
\begin{lstlisting}
Overall
1. Could you tell me about how you normally use generative AI in the animation pre-production process, and what your general thoughts are about using generative AI?
2. Please share your overall experience of using this system from beginning to end. What do you think is the biggest difference compared to the AI systems you've used before?
3. What do you find the most satisfying part of using this system, and why?
4. Which feature(s) of the system do you think are the most helpful, and why?
5. Did you encounter any challenges or frustrations while using the system?

Communication and Task Completion
1. Do you feel you could effectively communicate with the system so that it helped you complete tasks? How is this different from the baseline or your previous workflow?
2. In the pre-production tasks you carried out earlier, which included multiple stages and tasks, do you feel the system supported you well in coordinating and integrating across these tasks? How does this compare to the baseline or your previous workflow?
3. Do you feel the system helped you obtain the results you wanted more effectively? Why?
4. Do you think the results provided by the system were consistent?

Information Organization and Decision-Making
1. When collaborating with the system across a series of tasks, how do you feel the system helped you organize information? Did you experience information overload? How does this compare to your past workflow?
2. During the pre-production process, were you able to easily find and make use of the various kinds of information (previously generated content)? How does this differ from your past workflow?
3. Do you feel the system's design helps you understand the relationships between different stages and boards?
4. Do you think the system's design helped you make better creative decisions, such as selecting preferred designs or storylines?

Facilitating User Agency and Creative Exploration
1. Do you feel the system reduced tedious tasks and allowed you to focus more on the creative process? Why? Could you give an example?
2. When using the system, did you feel you were in control of the overall creative process? How is this different from your previous workflow?
3. Do you feel the system's results or suggestions helped you expand your creativity? Compared to your previous approach, what's different?
4. Both systems can expand creativity and provide different feedback -- roughly similar in this respect.

Overall Workflow and Creative Assistant
1. Do you think the system supported you well in developing from an initial rough idea to a more complete pre-production result?
2. From the perspective of a creative assistant, what kind of experience did you have with this system? Did it help you handle project workflow and reduce the burden of managing all kinds of information?
3. Do you have any other suggestions for the system, such as features you would like to see added?
4. If you were to use this system in future creative work, how would you use it? For example, if your job was scriptwriting, it could act as a discussion partner, so you wouldn't need everyone brainstorming at the same time, and it could help organize ideas.
\end{lstlisting}

\FloatBarrier
\section{Appendix C: Figures}
\label{Appendix_Figures}
\begin{figure*}[!t]
    \centering
    \includegraphics[width=1\linewidth]{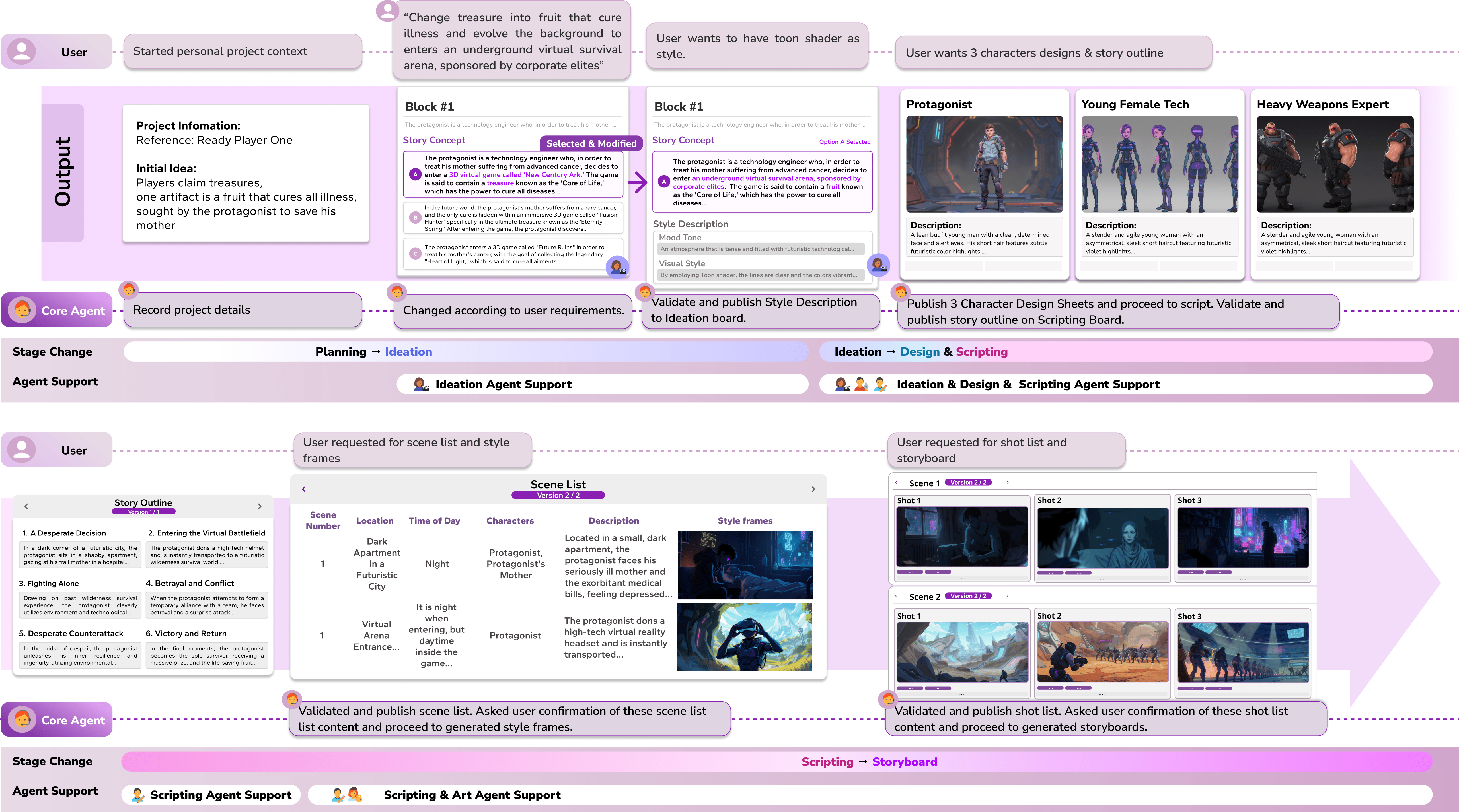}
    \caption{Detailed workflow and results of P8’s field study with AnimAgent. Beginning from the project context, he requested story concepts and modified the outputs, specifying a toon shader as the visual style. The workflow continued through character designs and generated scene content, and ultimately culminated in shot lists and complete storyboards.}
    \Description[This figure shows the workflow and results of a field study participant using AnimAgent, progressing from initial project context to storyboards.]{This figure shows the detailed workflow and results of participant P8’s field study with AnimAgent. The process begins with defining the project context, followed by requesting and modifying story concepts. The participant specified a toon shader as the preferred visual style. The workflow then advances 3 character designs, developing a story outline, and creating scene lists accompanied by styleframes. Finally, the process culminates in producing shot lists and complete storyboards. The figure illustrates how AnimAgent supports iterative refinement across stages, from initial concepts to final visual outputs.}
    \label{fig:field_study_detailed}
\end{figure*}


\end{document}